\documentclass[11pt, oneside]{article} 
\usepackage[margin=2cm]{geometry}                		
\usepackage[parfill]{parskip}    		
\usepackage{graphicx}			
\usepackage[utf8]{inputenc}
\usepackage[english]{babel} 
\usepackage{amsmath,amssymb,amsthm,bbm} 
\usepackage{mathrsfs}
\usepackage[margin=2cm]{geometry}
\usepackage[color=yellow]{todonotes}
\usepackage[colorlinks]{hyperref}
\usepackage{enumerate}

\usepackage{pgfplots}
\usepackage{float}
\usepackage{tikz,tikz-cd}
\usetikzlibrary{positioning}

\usepackage{graphicx}
\usepackage{caption}
\usepackage{multicol}
\usepackage{subcaption}
\usepackage{natbib}
\extrafloats{100}
\numberwithin{equation}{section}

\newtheorem{theorem}{Theorem}[section]

\newtheorem{remark}[theorem]{Remark}

\theoremstyle{definition}

\pgfplotsset{compat=1.16}

\def\contract{\makebox[1.2em][c]{\mbox{\rule{.6em}
{.01truein}\rule{.01truein}{.6em}}}}
\newcommand{\KN}{\mathbin{\bigcirc\mspace{-15mu}\wedge\mspace{3mu}}}

\begin{document}

\title{\textbf{Irreversible dynamics on Poisson manifolds}}
\author{Erwin Luesink \thanks{\href{mailto:e.luesink@uva.nl} 
{e.luesink@uva.nl} Korteweg-de Vries Institute of Mathematics, University of Amsterdam}
}
\date{}

\maketitle

\begin{abstract}
We present a geometric construction of irreversible dynamics on Poisson manifolds that satisfies the axioms of metriplectic mechanics and the GENERIC framework. Our approach relies solely on the underlying Poisson structure and its deformation theory, without requiring any additional metric structure. Specifically, we show that if the second Lichnerowicz–Poisson cohomology group of a Poisson manifold is nontrivial, one can construct a symmetric bracket that generates irreversible dynamics compatible with energy conservation and entropy production. This bracket is derived from a 2-cocycle that deforms the original Poisson structure, thereby modifying the associated Casimir foliation. We illustrate the construction with two finite-dimensional examples and one infinite-dimensional example: the duals of the Lie algebras of the special Euclidean group $SE(2)$, the Galilei group $SGal(3)$ and the group of diffeomorphisms over the circle. These examples demonstrate the applicability of the method in classical mechanics, control theory, and mathematical physics.
\end{abstract}

\section{Introduction}
Nonequilibrium thermodynamics has been an active field of research for over two centuries. Foundational contributions by Clausius, Kelvin, Maxwell, and Rayleigh laid the groundwork for a general theoretical understanding of irreversible processes. However, what is now known as the classical theory of nonequilibrium thermodynamics is primarily based on the seminal works \cite{onsager1931reciprocal, onsager1931reciprocal2}, in which the reciprocal relations connecting the linear coefficients in the phenomenological equations that relate thermodynamic fluxes and forces were introduced.

Over the last century, the field of nonequilibrium thermodynamics has diversified into a zoo of different approaches. Some treatments remain largely phenomenological, while others are rooted in more fundamental approaches using variational principles or geometric structures. A wide range of monographs that reflect these perspectives include \cite{biot1970variational, glansdorff1971thermodynamic, stueckelberg1971thermocinetique, woods1975thermodynamics, lavenda1978thermodynamics, truesdell1984rational, kondepudi2008modern, de2013non}. A modern variational theory of irreversible processes is developed in \cite{gay2017lagrangian, gay2017lagrangian2, gay2018lagrangian}, based on the Lagrange-d'Alembert variational principle. The port-Hamiltonian framework introduced by \cite{maschke1992port} is well-suited by design for modelling irreversible dynamics by means of the resistive ports that exist in this framework. Specifically, in the context of thermodynamics, a port-Hamiltonian approach is discussed in \cite{ramirez2022overview}. There are also modelling approaches to thermodynamics based on contact geometry, \cite{grmela2014contact, van2018geometry, bravetti2019contact}, where the irreversible part arises from the Reeb vector field associated with the contact form.

Several axiomatic frameworks have been proposed to model geometric irreversible mechanics and nonequilibrium thermodynamics. The first general description of dissipative dynamical systems dates back to \cite{willems1972dissipative}. Another formalism, introduced in \cite{kaufman1984dissipative} and generalised by \cite{morrison1986paradigm}, is metriplectic mechanics, which constructs irreversible dynamics by combining a Hamiltonian structure with a dissipative gradient-flow structure. The goal of metriplectic mechanics is to describe thermodynamically consistent systems, particularly in plasma physics and fluid dynamics. Another prominent framework is GENERIC (General Equation for Non-Equilibrium Reversible–Irreversible Coupling), introduced in \cite{grmela1997dynamics} and \cite{ottinger1997dynamics}. Interestingly, GENERIC and metriplectic mechanics share the same underlying axiomatic structure, despite arising from different communities and applications. 

A key challenge in the metriplectic and GENERIC approaches is to formulate an explicit symmetric structure that encodes irreversibility while preserving energy. Recently, in \cite{goreac2025generating}, irreversible dynamics is constructed based a geometric structure that assigns a dissipating vector field to an arbitrary Hamiltonian function and an arbitrary entropy function so that the resulting flow automatically satisfies the laws of thermodynamics. In this work, we provide an explicit construction of such a symmetric structure on Poisson manifolds---under certain conditions---that satisfies the axioms of irreversible mechanics based solely on the Poisson structure that underlies the reversible part. Notably, our construction does not rely on any additional Riemannian structure, the symmetric structure can be constructed exclusively from Poisson geometric notions.

\paragraph{Main result.} Our main result can be stated as follows: Given a Poisson manifold $(M,\{\,\cdot\,,\,\cdot\,\})$ with a degenerate Poisson structure $\pi$ that has at least one Casimir, if the second Lichnerowicz-Poisson cohomology $H^2_\pi(M)$ group is nontrivial, then $M$ admits irreversible dynamics that satisfy the axioms of metriplectic mechanics and GENERIC. 

There has to be at least one Casimir, since the functionally independent Casimirs are entropies in the metriplectic and GENERIC theories. The nontriviality of the second Lichnerowicz–Poisson cohomology guarantees the existence of topologically nontrivial deformations of the Poisson bracket. In particular, this implies that the Casimir foliation associated with the original Poisson structure can be altered by such a deformation. Consequently, a function that was a Casimir for the original Poisson bracket may no longer be a Casimir for the deformed bracket. Using this fact, one can explicitly construct a dissipative structure from the tensor product of the 2-cocycle responsible for the deformation. This structure encodes irreversible dynamics consistent with the underlying geometric framework. In other words, we can guarantee energy preservation, while increasing (or decreasing, depending on convention) the entropy, in agreement with the second law of thermodynamics.

We provide two detailed finite-dimensional examples of irreversible dynamics constructed using this approach. The first example discusses two constructions of irreversible dynamics on the special Euclidean group $SE(2)$, which is encountered in planar robotics and control theory, in physics and chemistry (e.g. the modelling of rigid molecules constrained to a plane), and in computer vision and image processing. The second example concerns the Galilei group, which describes transformations between inertial frames and serves as the symmetry group of Newtonian mechanics.

\paragraph{Outlook.} The approach that we present also works in the infinite-dimensional case. In fact, the necessary condition for our construction (the nontriviality of the second Lichnerowicz-Poisson cohomology group) is almost always satisfied for models encountered in continuum mechanics. Detailed examples of infinite-dimensional irreversible dynamics constructed using the approach in the present work are the topic of future works.

\paragraph{Overview of the paper.} In Section \ref{sec:prelims}, we review the mathematical preliminaries required to discuss dynamics on Poisson manifolds. We introduce the graded algebra of multivector fields and the Schouten–Nijenhuis bracket. We further discuss the Lichnerowicz–Poisson cochain complex, with particular focus on its second cohomology group. In Section \ref{sec:irreversiblemech}, we present the axioms of metriplectic mechanics and the GENERIC framework, and we describe the construction of the symmetric bracket governing the irreversible dynamics. Section \ref{sec:irreversiblese2} is devoted to the construction of irreversible dynamics on $\mathfrak{se}(2)^*$, the dual of the Lie algebra $\mathfrak{se}(2)$ corresponding to the special Euclidean group $SE(2)$. We illustrate the kind of dynamics the framework leads to by considering the example of Kirchhoff equations for a 2D rigid body moving in an ideal fluid. In Section \ref{sec:irreversiblegal}, we construct irreversible dynamics on $\mathfrak{sgal}(3)^*$, the dual of the Galilei Lie algebra $\mathfrak{sgal}(3)$, associated with the Lie group $SGal(3)$. This example reveals several striking parallels with familiar structures in quantum mechanics. In Section \ref{sec:irrevvects1}, we discuss irreversible dynamics on the infinite-dimensional group of diffeomorphisms of the circle and show that one obtains the Korteweg-de Vries equation with a nonlocal dispersion coefficient. Finally, in Section \ref{sec:conclusion}, we provide a discussion and some concluding remarks.

\section{Poisson geometry preliminaries}\label{sec:prelims} 
In this section, we recall the necessary geometric, topological and algebraic preliminaries for setting up Hamiltonian mechanics on Poisson manifolds. For more background on geometric mechanics, we refer to \cite{abraham1978foundations, holm2009geometric, marsden2013introduction}. We closely follow \cite{crainic2021lectures} for the necessary tools of Poisson geometry. 

A Poisson manifold is a smooth manifold $(M, \{\,\cdot\,,\,\cdot\,\})$ endowed with a Lie bracket on the space $C^\infty(M)$ of smooth functions
\begin{equation}
    \{\,\cdot\,,\,\cdot\,\} : C^\infty(M) \times C^\infty(M) \to C^\infty(M),
\end{equation}
which also satisfies the Leibniz rule. For any $f,g,h\in C^\infty(M)$:
\begin{align}
	\{f, gh\} = \{f, g\}h + g\{f, h\}.
\end{align}
A Poisson map between Poisson manifolds $(M_1,\{\,\cdot,,\,\cdot\,\}_1)$ and $(M_2,\{\,\cdot,,\,\cdot\,\}_2)$ is a smooth map $\Phi:M_1\to M_2$ which induces a Lie algebra homomorphism 
\begin{equation}
	\{f\circ\Phi,g\circ\Phi\}_1 = \{f,g\}_2\circ \Phi,
\end{equation}
for all $f,g\in C^\infty(M_2)$. A Poisson bracket can be degenerate. The kernel of a Poisson bracket is spanned by the Casimir functions \( C \in C^\infty(M) \), defined by the condition
\begin{equation}\label{eq:casimirs}
    \{C, f\} = 0 \quad \text{for all } f \in C^\infty(M).
\end{equation}
Casimirs are preserved under every Hamiltonian flow and remain constant along the symplectic leaves of the foliation determined by the Poisson structure. They play a fundamental role in the geometry and integrability of Hamiltonian systems on Poisson manifolds. 

In a local chart $(U,x^1,\hdots, x^m)$, a Poisson bracket can be expressed in terms of smooth functions $\pi^{ij}\in C^\infty(U)$
\begin{equation}
\{f,g\}|_U = \sum_{i,j}^m \pi^{ij}\frac{\partial f}{\partial x^i}\frac{\partial g}{\partial x^j}.
\end{equation} 
The bivector $\pi=\sum
_{i<j}\pi^{ij}\frac{\partial}{\partial x^i}\wedge\frac{\partial}{\partial x^j}$ thus gives a Poisson bracket under certain conditions. Bilinearity, skew-symmetry, and the Leibniz rule follow directly from the definition $\{f, g\} = \pi(df, dg) $. However, the Jacobi identity is more subtle. It holds if and only if the bivector $\pi$ satisfies the condition
\begin{equation}\label{eq:schoutennijenhuis}
[\pi, \pi] = 0,
\end{equation}
where \( [\,\cdot\,,\,\cdot\,] \) is the Schouten--Nijenhuis bracket. Before introducing the Schouten--Nijenhuis bracket in a detailed manner, let us give some practical consequences of this necessary and sufficient condition. Equation \eqref{eq:schoutennijenhuis} is a nonlinear partial differential equation that is, in general, difficult to solve explicitly. For this reason, it is often more practical to define a Poisson manifold via a Poisson bracket satisfying the Jacobi identity and then recover the associated bivector field, rather than starting with a bivector field and checking whether it defines a Poisson structure. 

The Poisson bivector is a multivector field of degree 2. In general, a multivector field $\vartheta$ of degree $k$ on $M$ is an element of the space $\mathfrak{X}^k(M):=\Gamma(\bigwedge^k TM)$ and can be identified with a $C^\infty(M)$-multilinear, alternating map of degree $k$ on the space $\Omega^1(M)$ of 1-forms on $M$. Under this identification, analogous to constructions on differential forms, one can describe a wedge product to introduce a grading
\begin{equation}
    \wedge:\mathfrak{X}^k(M)\times\mathfrak{X}^l(M)\to\mathfrak{X}^{k+l}(M)
\end{equation}
by the formula
\begin{equation}
    (\vartheta\wedge\zeta)(\alpha_1,\hdots,\alpha_{k+l}) = \sum_{\sigma\in S_{k,l}}(-1)^\sigma\vartheta(\alpha_{\sigma(1)},\hdots, \alpha_{\sigma(k)})\zeta(\alpha_{\sigma(k+1)},\hdots, \alpha_{\sigma(k+l)}),
\end{equation}
with the sum over all $(k,l)$-shuffles, i.e., permutations $\sigma\in S_{k+l}$ such that $\sigma(1)<\cdots<\sigma(k)$ and $\sigma(k+1)<\cdots <\sigma(k+l)$. For $k=0$, the space $\mathfrak{X}^0(M)=C^\infty(M)$, i.e., the ring over which these modules are defined. The wedge product on multivector fields is graded commutative and associative
\begin{equation}
    \begin{aligned}
    \vartheta\wedge\zeta &= (-1)^{\deg\vartheta\deg\zeta}\zeta\wedge\vartheta,\\
    (\vartheta\wedge\zeta)\wedge\tau &= \vartheta\wedge(\zeta\wedge\tau).
    \end{aligned}
\end{equation}
The grading for multivector fields is by total degree, that is, $\deg \theta = k$ if $\theta\in \mathfrak{X}^k(M)$. The properties of the wedge product give
\begin{equation}
    \mathfrak{X}^\bullet(M) = \bigoplus_{k=0}^m\mathfrak{X}^k(M)
\end{equation}
the structure of a graded commutative algebra. On $\mathfrak{X}^\bullet(M)$, the Schouten-Nijenhuis bracket is the unique $\mathbb{R}$-bilinear operation $[\,\cdot\,,\,\cdot\,]:\mathfrak{X}^{k+1}(M)\times\mathfrak{X}^{l+1}(M)\to\mathfrak{X}^{k+l+1}(M)$ with $k,l\geq-1$ with the following properties
\begin{enumerate}
    \item When $k=l=0$, the Schouten-Nijenhuis bracket is the usual Lie bracket of vector fields.
    \item When $k=0$ and $l=-1$, it is the Lie derivative: $[X,f] = \mathcal{L}_X f = X(f).$
    \item It is graded skew-symmetric: $[\vartheta,\zeta] = -(-1)^{kl}[\zeta,\vartheta]$, for $\vartheta\in\mathfrak{X}^{k+1}(M)$ and $\zeta\in\mathfrak{X}^{l+1}(M)$.
    \item It satisfies the graded Leibniz identity: $[\vartheta,\zeta\wedge\tau] = [\vartheta,\zeta]\wedge\tau + (-1)^{k(l+1)}\zeta\wedge[\vartheta,\tau]$, for $\vartheta\in\mathfrak{X}^{k+1}(M), \zeta\in\mathfrak{X}^{l+1}(M)$ and $\tau\in\mathfrak{X}^{m+1}(M)$.
\end{enumerate}
The Schouten-Nijenhuis bracket also satisfies the graded Jacobi identity
\begin{equation}
    (-1)^{km}[\vartheta,[\zeta,\tau]]+(-1)^{lk}[\zeta,[\tau,\vartheta]] + (-1)^{lm}[\tau,[\vartheta,\zeta]] = 0,
\end{equation}
for $\vartheta\in\mathfrak{X}^{k+1}(M), \zeta\in\mathfrak{X}^{l+1}(M)$ and $\tau\in\mathfrak{X}^{m+1}(M)$. The properties of the Schouten-Nijenhuis bracket mean that the space $(\mathfrak{X}^\bullet(M),[\,\cdot\,,\,\cdot\,])$ has the structure of a Gerstenhaber algebra, which plays a central role in Poisson geometry and deformation quantisation, because its cohomology theory determines when a Poisson bracket can be deformed into another Poisson bracket in a nontrivial way. 

Given a bivector field $\pi \in \mathfrak{X}^2(M)$, one defines the Lichnerowicz--Poisson differential $d_\pi$ on multivector fields by
\begin{equation}
    d_\pi \vartheta := [\pi, \vartheta],
\end{equation}
for any $\vartheta \in \mathfrak{X}^\bullet(M)$, where $[\,\cdot\,,\,\cdot\,]$ denotes the Schouten--Nijenhuis bracket. The bracket properties ensure that $d_\pi$ is a graded derivation of degree $+1$, and it satisfies $d_\pi^2 = 0$ if and only if $[\pi, \pi] = 0 $. This can be seen in the following way. Suppose that $[\pi,\pi]=0$, then it follows from the graded Jacobi identity of the Schouten-Nijenhuis bracket that $d_{\pi}^2:= d_{\pi}\circ d_{\pi} =[\pi,[\pi,\,\cdot\,]] = \frac{1}{2}[[\pi,\pi],\,\cdot\,]= 0$. In this case, the bivector $\pi$ is a Poisson bivector, and the operation $d_\pi$ defines a differential graded algebra structure on $\mathfrak{X}^\bullet(M)$. The resulting cochain complex $(\mathfrak{X}^\bullet(M),d_\pi)$
\begin{equation}
    \cdots \longrightarrow \mathfrak{X}^{k-1}(M) \xrightarrow{d_\pi} \mathfrak{X}^k(M) \xrightarrow{d_\pi} \mathfrak{X}^{k+1}(M) \longrightarrow \cdots
\end{equation}
is the Lichnerowicz--Poisson complex, and its cohomology groups
\begin{equation}
    H^k_\pi(M) := \frac{\ker(d_\pi : \mathfrak{X}^k(M) \to \mathfrak{X}^{k+1}(M))}{\operatorname{im}(d_\pi : \mathfrak{X}^{k-1}(M) \to \mathfrak{X}^k(M))}
\end{equation}
are the Lichnerowicz-Poisson cohomology groups of $(M, \{\,\cdot\,,\,\cdot\,\})$. A Lichnerowicz-Poisson $k$-cocycle is an object $\theta$ such that $d_\pi\theta = 0$. The space of cocycles is denoted $Z_\pi^k(M)$ and the space of $k$-coboundaries is denoted $B^k_\pi(M)$. The cohomology groups $H^k_\pi(M) \simeq Z^k_\pi(M)/B^k_\pi(M)$ consist of $k$-cocycles that are not $k$-coboundaries. These cohomology groups play a central role in the deformation theory of Poisson structures. A $2$-coboundary $a$ is a $2$-cocycle that is obtained by transporting a Poisson structure along a vector field $X$, i.e. $a=[\pi, X]$. The second Lichnerowicz-Poisson cohomology group $H^2_\pi(M)$ classifies infinitesimal deformations modulo trivial ones. The second Lichnerowicz-Poisson cohomology is particularly important for the development of an irreversible formulation of mechanics. 

Any higher order Poisson structure beyond the constant structure may have a nontrivial kernel. The kernel is important, since it provides candidates for the entropy function in geometric dissipative mechanics, as we describe in the next section.

In this section, we discussed the mechanical preliminaries. In the next section, we discuss the axioms of metriplectic mechanics and GENERIC, and focus on explicitly obtaining the symmetric bracket that defines the irreversible part of the dynamics.

\section{Irreversible energy-preserving mechanics}\label{sec:irreversiblemech}
An irreversible energy-preserving system consists of a Poisson manifold $(M,\{\,\cdot\,,\,\cdot\,\})$, a Poisson structure $\pi\in\Gamma(\bigwedge^2TM)$, a symmetric structure $\chi\in\Gamma(S^2 TM)$ (a symmetric $(2,0)$-tensor), and two functions $H,S\in C^\infty(M)$. The function $H$ is the Hamiltonian, describing the energy in the system, and the function $S$ is the entropy. Axiomatically, irreversible energy-preserving dynamics is then expressed in terms of two binary brackets on functions $f,g\in C^\infty(M)$. The first bracket is the Poisson bracket $\{\,\cdot\,,\,\cdot\,\}:C^\infty(M)\times C^\infty(M)\to C^\infty(M)$, associated with the Poisson bivector $\pi$
\begin{equation}
    \{f,g\} = \pi(df,dg).
\end{equation}
The second bracket is a symmetric bracket $((\,\cdot\,,\,\cdot\,)):C^\infty(M)\times C^\infty(M)\to C^\infty(M)$, defined as
\begin{equation}
((f,g)):= \chi(df,dg),
\end{equation}
and is a positive semidefinite symmetric bracket, i.e., $((\,\cdot\,,\,\cdot\,))$ is bilinear and symmetric, and $((f,f))\geq 0$ for every $f\in C^\infty(M)$. To obtain irreversible energy-preserving dynamics, we further require the following two conditions
\begin{equation}\label{eq:kernelconditions}    
\{S,f\} = 0 \text{ and } ((H,f))= 0 \text{ for all } f\in C^\infty(M),
\end{equation}
or equivalently, $\pi^\sharp(dS)=0$ and $\chi^\sharp(dH) = 0$. The practical implications of the first condition is that the entropy is a Casimir of the Poisson bracket and the second condition ensures that entropy is non-decreasing (in agreement with the second law of thermodynamics) while preserving the energy. It is clear that from these axioms, a nondegenerate Poisson bivector cannot be used since there would be no nontrivial entropy function.

Irreversible energy-preserving dynamics can then be expressed using a generator $\mathcal{F} = H + \tau S$, where $H$ is the Hamiltonian, $S$ is the entropy and $\tau$ can be interpreted as a global constant temperature. The physical interpretation of $\mathcal{F}$ is a Gibbs free energy function, modulo a sign convention. Given any observable $f\in C^\infty(M)$, irreversible dynamics is then expressed, using the compatibility properties, as
\begin{equation}
    \begin{aligned}
        \frac{\boldsymbol{d}}{\boldsymbol{d}t}f &= \{\mathcal{F},f\} + ((\mathcal{F},f)) \\
        &= \{H+\tau S,f\} + ((H+\tau S,f))\\
        &= \{H,f\} + \tau((S,f)).
    \end{aligned}
\end{equation}
It is common to set, without loss of generality, $\tau=1$. Given the properties of the Poisson bracket and the symmetric bracket, we see that 
\begin{equation}
    \begin{aligned}
        \frac{\boldsymbol{d}}{\boldsymbol{d}t}H &= 0,\\
        \frac{\boldsymbol{d}}{\boldsymbol{d}t}S &= ((S,S))\geq 0 .
    \end{aligned}
\end{equation}
For many conservative systems, such as ideal fluid dynamics, rigid bodies, ideal magnetohydrodynamics, the Hamiltonian formulation is known. This means that the Poisson bracket as well as the Hamiltonian are known. The key question for irreversible formulations of mechanics is then how to obtain the symmetric structure $\chi$, or equivalently, how to obtain the symmetric bracket $((\,\cdot\,,\,\cdot\,))$ such that the above properties hold, and to determine a suitable entropy. Since the entropy is required to be a Casimir, the Poisson structure determines the family of candidates for the entropy. 

\paragraph{Kulkarni-Nomizu product.}
In \cite{morrison2024inclusive}, it is explained that one way to construct the symmetric bracket is to first obtain a contravariant $(4,0)$-tensor and use this to induce the symmetric bracket. In order for such a tensor to be able to induce the symmetric bracket, it has to have certain symmetry properties. We define a 4-bracket on functions $f,g,h,k\in C^\infty(M)$ in terms a of $(4,0)$-tensor $T:\Omega^1(M)^{\times 4}\to C^\infty(M)$
\begin{equation}\label{eq:4bracket}
    (f,g; h,k) := T(df,dg,dh,dk).
\end{equation}
In order to satisfy the axioms of irreversible mechanics, the $4$-bracket and thereby the $(4,0)$-tensor $T$, should have properties 1,2 and 3 from the following list of algebraic symmetries
\begin{enumerate}
    \item $(f,g;h,k) = -(g,f;h,k)$.
    \item $(f,g;h,k) = -(f,g;k,h)$.
    \item $(f,g;h,k) = (h,k;f,g)$.
    \item $(f,g;h,k) + (g,h;f,k) + (h,f;g,k)=0.$
\end{enumerate}
Note that the semicolon notation is used to emphasize the symmetry between the pairs $(f,g)$ and $(h,k)$. Together with property 4, these are precisely the algebraic symmetries of the algebraic curvature tensors in Riemannian geometry. The fourth property is the algebraic Bianchi identity, which in Riemannian geometry links the curvature tensor with a torsion-free connection. From the 4-bracket and the Hamiltonian, one obtains a symmetric bracket
\begin{equation}\label{eq:symbracket}
    ((f,g)) := (f,H;g,H) = T(df,dH,dg,dH)
\end{equation}
which satisfies
\begin{equation}
    \begin{aligned}
        ((f,g)) &= (f,H;g,H) = (g,H;f,H) = ((g,f)),\\
        ((H,f)) &= (H,H;f,H) = -(H,H;f,H) = 0,
    \end{aligned}
\end{equation}
so we obtain the desired algebraic properties. For consistency with the second law of thermodynamics, it is further required that the 4-bracket satisfies $T(df,dg,df,dg) = (f,g;f,g) \geq 0$ for all $f,g\in C^\infty(M)$ (which ensures $((f,f)) \geq 0$). This is a property that does not follow from the algebraic symmetries.

At this stage, we have established the following. To obtain a symmetric $2$-bracket that satisfies the axioms of metriplectic mechanics, it is sufficient to find a $4$-bracket that satisfies three algebraic symmetries. We defined such a $4$-bracket in terms of a contravariant $(4,0)$-tensor $T$ and noted that this tensor must have the same algebraic symmetry properties. So the next step is to construct such a $(4,0)$-tensor $T$. 

To obtain such a $(4,0)$-tensor explicitly, one suggestion by \cite{morrison2024inclusive} is to use two symmetric $(2,0)$-tensors and compute their Kulkarni-Nomizu (KN) product. This product was introduced for the construction of algebraic curvature tensors in Riemannian geometry, see \cite{lee2018introduction}, but has also appeared in the context of Nambu mechanics in \cite{alekseevsky96decomposability}. The KN product of two symmetric covariant $(0,2)$-tensors produces a covariant $(0,4)$-tensor that has the algebraic symmetries of a curvature tensor. Hence, a contravariant version of the KN product is a natural choice for the construction of the $(4,0)$-tensor $T$, since one guarantees the list of algebraic symmetries. Given two symmetric $(2,0)$-tensors, one can build a $(4,0)$-tensor with the same symmetries as the covariant analogue using a contravariant version of the KN product. Given two $(2,0)$-tensors $\sigma,\mu$, one can construct the $(4,0)$-tensor $T$ via
\begin{equation}
    \begin{aligned}
    T(df,dg,dh,dk) :&= (\sigma\KN \mu)(df,dg,dh,dk)\\
    :&= \sigma(df,dh)\mu(dg,dk) + \sigma(dg,dk)\mu(df,dh)\\
    &\quad - \sigma(df,dk)\mu(dg,dh) - \sigma(dg,dh)\mu(df,dk)
    \end{aligned}
\end{equation}
where the second line is the definition of the KN product, though here it is applied to symmetric contravariant 2-tensors, rather than symmetric covariant 2-tensors for which this product was originally defined. A $(4,0)$-tensor constructed using the KN product satisfies
\begin{equation}
    \begin{aligned}
    (\sigma\KN\mu)(df,dg,df,dg) &= \sigma(df, df)\mu(dg, dg) + \sigma(dg, dg)\mu(df, df) - 2\sigma(df, dg)\mu(df, dg)
    \end{aligned}
\end{equation}
In case the tensors used in the KN product are positive definite, we can show that the resulting object is also positive definite. By the arithmetic mean--geometric mean inequality and the Cauchy-Schwarz inequality, we obtain
\begin{equation}
\begin{aligned}
\sigma(df,df)\mu(dg,dg) + \sigma(dg,dg)\mu(df,df) &\geq 2\sqrt{\sigma(df,df)\mu(dg,dg)\sigma(dg,dg)\mu(df,df)}\\
&\geq 2\sqrt{(\sigma(df,dg)^2\mu(df,dg)^2}\\
&= 2|\sigma(df,dg)\mu(df,dg)| \\
&\geq 2\sigma(df,dg)\mu(df,dg).
\end{aligned}
\end{equation}
Hence the KN product of two positive definite tensors leads to a $(4,0)$-tensor with the desired properties. 

\paragraph{Poisson geometry construction.}
We now provide an alternative approach that does not assume additional Riemannian structure and relies only on the Poisson structure. The tensor product of most bivectors with themselves is a valid choice for the construction of the contravariant tensor $T$, since this satisfies the first three properties by construction. To allow for an increase in entropy, it is required that the entropy is not in the kernel of this bivector. This rules out the use of Poisson bivector associated with the bracket. Indeed, given the Poisson structure on $M$, we have the Poisson bivector $\pi$. This is a skew-symmetric $(2,0)$-tensor. The tensor product $T:=\pi\otimes\pi$ is a $(4,0)$-tensor that by construction satisfies the algebraic symmetries of a contravariant algebraic curvature tensor with the exception of the contravariant version of the first Bianchi identity. Indeed, 
\begin{equation}
    T(df,dg,dh,dk) := \pi(df,dg)\pi(dh,dk)
\end{equation}
from which it can be checked that
\begin{equation}
    \begin{aligned}
    T(df,dg,dh,dk) &=  -T(dg,df,dh,dk),\\
    T(df,dg,dh,dk) &= -T(df,dg,dk,dh),\\
    T(df,dg,dh,dk) &=  T(dh,dk,df,dg),
    \end{aligned}
\end{equation}
which imply that the 4-bracket obtained via $T=\pi\otimes\pi$ in \eqref{eq:4bracket} satisfies the first three algebraic conditions. We also see that
\begin{equation}\label{eq:entropyinc}
    T(df,dg,df,dg) = \pi(df,dg)^2 \geq 0, \text{ for } f,g\in C^\infty(M).
\end{equation}
Of course, in the irreversible setting, one takes the entropy $S$ to be a Casimir of $\pi$. This also means that the tensor product of the Poisson structure is not directly suitable for the symmetric bracket, since it would vanish on $S$, but it gives an indication of which structure would be suitable. While not suitable for irreversible dynamics following the axioms of metriplectic dynamics and GENERIC, the tensor product construction $T=\pi\otimes\pi$ involving the Poisson bivector $\pi$ is implicitly used already by \cite{brockett1991dynamical}, and can be interpreted as double-bracket dissipation, which dissipates the Hamiltonian. If one selects a different Hamiltonian (say $H_1$) in the symmetric bracket \eqref{eq:symbracket} compared to the Hamiltonian (say $H_0$) used for the conservative dynamics, the Hamiltonian $H_0$ will be dissipated. In \cite{luesink2025symplectic}, double-bracket dissipation was used to construct Langevin dynamics on reductive Lie groups. This Langevin dynamics can be formulated fully in terms of Poisson brackets and the dissipative term is of the tensor product form. 

Given a Poisson manifold, the Poisson structure generating the bracket is itself not the correct object to take the tensor product with itself of for irreversible dynamics because of its kernel. In the next section, we discuss how to address this problem and use the tensor product construction as above.

\subsection{Compatible Poisson structures and generic 2-cocycles}
To have an irreversible formulation in which the entropy is able to increase, we need to introduce an anti-symmetric object on the Poisson manifold that does not have the chosen entropy as a Casimir. Modifying Casimirs changes the foliation of the manifold into symplectic leaves. We distinguish regular symplectic leaves from singular symplectic leaves and consider level sets of Casimirs that give rise to the regular symplectic leaves to construct an irreversible and energy-conserving theory.

As the starting point, we consider a deformation of a Poisson bracket. In case we insist on perturbing a Poisson bracket with a 2-cocycle to obtain another Poisson bracket, we take $\epsilon\in \mathbb{R}$ and define $\pi_\epsilon$ as a deformation of $\pi$ in the following way
\begin{equation}
    \pi_\epsilon = \pi + \epsilon a,
\end{equation}
where $a$ is a skew-symmetric contravariant $(2,0)$-tensor. For $\pi_\epsilon$ to generate a Poisson bracket, it must be bilinear, skew-symmetric and satisfy the Schouten-Nijenhuis equation \eqref{eq:schoutennijenhuis}. Bilinearity and skew-symmetry follow directly by construction. To satisfy the Schouten-Nijenhuis equation, we compute
\begin{equation}
    \begin{aligned}
        0 &= [\pi_\epsilon, \pi_\epsilon] \\
        &= [\pi+\epsilon a,\pi+\epsilon a]\\
        &= [\pi,\pi] + 2\epsilon[\pi,a] + \epsilon^2[a,a].
    \end{aligned}
\end{equation}
Since $\pi$ is associated with a Poisson bracket, $[\pi,\pi]=0$. We thus obtain the following constraints on $a$
\begin{equation}
    [\pi,a]=0 \text{ and } [a,a]=0.
\end{equation}
The first constraint implies that $a$ is a Poisson 2-cocycle in the Lichnerowicz-Poisson cohomology defined by $\pi$. This means that $a$ lies in the kernel of the Poisson differential $d_{\pi}=[\pi,\,\cdot\,]$, defined using the Schouten-Nijenhuis bracket acting on the graded space $\mathfrak{X}^\bullet (M)$ of multivector fields. The second constraint implies that $a$ is itself a Poisson bivector.

It is not necessary for the present construction that the condition $[a,a]=0$ holds, i.e., $a\otimes a$ is a suitable candidate for generating the symmetric bracket associated with irreversible dynamics whenever the condition $d_\pi(a) = [\pi, a]=0$ is holds, with $a$ being a Poisson 2-cocycle that is not a 2-coboundary. To understand why a 2-coboundary fails to generate physically meaningful dissipation, let $X \in \mathfrak{X}(M)$ be a vector field on the Poisson manifold. The associated 2-coboundary $a$ in the Lichnerowicz-Poisson complex is given by the negative Lie derivative, $a = - \mathcal{L}_X \pi$. We evaluate this coboundary on the gradients of the entropy $S$ and the Hamiltonian $H$. Using the Leibniz rule for the Lie derivative, we obtain
\begin{equation}
	a(dS, dH) = X(\{S, H\}) - \{X(S), H\} - \{S, X(H)\}.
\end{equation}
The entropy $S$ is a Casimir, so we have $\{S, H\} = 0$ and $\{S, X(H)\} = 0$ and the expression simplifies to	$a(dS, dH) = -\{X(S), H\}$. The resulting entropy production in our framework would be $((S, S)) = \{X(S), H\}^2$. Mathematically, it is possible for this term to be positive, e.g., if one chooses a vector field $X$ that is everywhere transverse to the symplectic leaves. However, the entropy $S$ is not an independent generator of motion, since only $X(S)$ plays a role in the dissipative dynamics. The bracket $\{X(S), H\} = X_H(X(S))$ represents the rate of change of the function $X(S)$ strictly along the conservative Hamiltonian vector field $X_H$. Since the supposed dissipation is entirely slaved to the reversible dynamics, the entropy $S$ fails to act as an independent generator of motion. Indeed, if the conservative flow halts, i.e., when $X_H = 0$, the entropy production identically vanishes. 

From a geometric point of view, a symmetric bracket built from a 2-coboundary represents a continuously shifting coordinate gauge rather than true irreversible relaxation. The symplectic leaves are simply being dragged along the flow of $X$, creating a coordinate artifact rather than true thermodynamic relaxation. Hence, constructing a physically meaningful symmetric bracket that drives true irreversible relaxation requires a nontrivial 2-cocycle. Thus, the quotient space
\begin{equation}
H^2_{\pi}(M) = \frac{\ker d_\pi \cap \Gamma\big(\bigwedge^2 TM\big)}{\operatorname{im} d_\pi \cap \Gamma\big(\bigwedge^2 TM\big)},
\end{equation}
is exactly the space of bivectors that admit genuine entropy production. A natural choice for a bivector with the required properties is therefore an element of the second Lichnerowicz-Poisson cohomology of the Poisson structure $\pi$. We provide an example of a construction using a 2-cocycle that does not satisfy the Schouten-Nijenhuis equation in Section \ref{sec:irreversiblese2}. 

Systems admitting two different Hamiltonian formulations are relatively common. In the theory of integrable systems, one encounters bi-Hamiltonian systems, which are Hamiltonian systems whose dynamics can be described by two Poisson structures $\pi_1,\pi_2$ that are not diffeomorphisms of one another, but are compatible in the sense that $[\pi_1,\pi_2]=0$. This is the core of the Lenard-Magri scheme which is often employed to prove that a dynamical system is completely integrable, \cite{magri1978simple, lenard1979recursion}. The compatibility means that each Poisson bivector is a 2-cocycle with respect to the Lichnerowicz-Poisson differential defined by the other, that is, $\pi_2\in Z^2_{\pi_1}$ and $\pi_1\in Z^2_{\pi_2}$. If the Casimir foliations of $\pi_1$ and $\pi_2$ differ, that is, if their kernels define distinct symplectic decompositions of the phase space, then the system admits a natural splitting into a Hamiltonian (conservative) component and a dissipative component, the latter constructed from the tensor product of the second Poisson bivector. Since differing Casimir foliations are necessary for such a conservative and dissipative splitting, the Mishchenko–Fomenko argument shift construction, illustrated in \cite{mishchenko1978generalized}, which defines $\pi_2$ as a linearisation of $\pi_1$ at a fixed point $a\in\mathfrak{g}^*$, is not suitable for irreversible dynamics, the resulting Poisson structures always share the same set of Casimir functions, and hence define the same foliation of the phase space.

A summary of the results so far in the present section is the following. The construction of a $(4,0)$-tensor that satisfies certain algebraic symmetries is sufficient for the construction of a symmetric bracket. The KN product is able to generate such a $(4,0)$-tensor, but one requires additional symmetric $(2,0)$-tensors to use this product. As an alternative, we construct the $(4,0)$-tensor by taking the tensor product of a nontrivial Poisson 2-cocycle with itself. This the Poisson geometric framework for irreversible dynamics, in which one structure governs the preservation of energy, while the other encodes entropy production or decay of Casimirs, yielding a dynamical system consistent with thermodynamic principles.

\begin{remark}
The approaches discussed in this section construct explicit symmetric brackets that satisfy the kernel conditions \eqref{eq:kernelconditions}. These degenerate symmetric brackets are associated with the degenerate $M$-operators in the GENERIC formulation, see \cite[Page 6623]{ottinger1997dynamics}. Both the approach using the KN product of \cite{morrison2024inclusive} as well as the present approach based on Poisson 2-cocycles construct symmetric brackets that depend on gradients of the Hamiltonian to guarantee that the kernel conditions are satisfied.  

This Hamiltonian dependence is unavoidable for a universal geometric construction. If the symmetric bracket and $M$-operator where purely state-dependent, this will restrict the class of admissible Hamiltonians. By explicitly embedding the Hamiltonian gradients into the symmetric bracket, the framework guarantees energy conservation for an arbitrary choice of Hamiltonian.
\end{remark}

We now ask under what conditions a generic Poisson manifold $(M, \pi)$ admits a nontrivial class in the second Lichnerowicz--Poisson cohomology group $H^2_{\pi}(M)$. Such a class is represented by bivector fields $a$ satisfying $d_\pi(a) = [\pi, a] = 0$ for which no vector field $X$ exists such that $a = -\mathcal{L}_X \pi = [\pi, X] = d_\pi(X)$.

In case the Poisson manifold arises from a Lie--Poisson structure on $\mathfrak{g}^*$, the dual of a finite-dimensional Lie algebra $\mathfrak{g}$, one can use Lie algebra cohomology to obtain explicit conditions under which $H^2_{\pi}(M)$ is nontrivial. There are subtle relations between Lichnerowicz--Poisson cohomology $H^\bullet_{\pi}(\mathfrak{g}^*)$ and the Chevalley--Eilenberg cohomology $H^\bullet_{CE}(\mathfrak{g})$ with real coefficients, for the details we refer to \cite{dufour2006poisson}. For our purposes, the important statement is the following, (Proposition 2.29 in \cite{crainic2021lectures}): There is a 1-to-1 correspondence between affine Lie-Poisson structures on $\mathfrak{g}^*$ and Lie algebra structures plus a 2-cocycle on $\mathfrak{g}$. In the finite-dimensional setting, by the Whitehead lemmas, if $\mathfrak{g}$ is semisimple, then $H^1_{CE}(\mathfrak{g},\mathbb{R}) = H^2_{CE}(\mathfrak{g},\mathbb{R}) = 0 $, see \cite{jacobson1979lie}. Hence, for finite-dimensional Lie algebras $\mathfrak{g}$, a necessary (but not sufficient) condition for the Lichnerowicz–Poisson cohomology $H^2_{\pi}(M)$ to be nontrivial in the case where $M = \mathfrak{g}^*$ with its Lie–Poisson structure is that $\mathfrak{g}$ is not semisimple. In particular, solvable or nilpotent Lie algebras can have nontrivial second Chevalley–Eilenberg cohomology, which leads to genuine deformations of the corresponding Lie-Poisson structure on $\mathfrak{g}^*$. 

Central extensions of Lie algebras are classified by $H^2_{CE}(\mathfrak{g},\mathbb{R})$ and an element of this group is a Lie algebra 2-cocycle. Upon translating this algebraic object onto the dual space, it manifests itself as a constant bivector. This means that the Schouten-Nijenhuis equation is trivially satisfied and thus central extensions always give deformations of Poisson brackets that are themselves Poisson brackets. In Section \ref{sec:irreversiblese2}, we give an example of irreversible dynamics on $\mathfrak{se}(2)^*$, the dual of the Lie algebra of the special Euclidean group, generated from a central extension, and a second example using a generic Poisson 2-cocycle. In Section \ref{sec:irreversiblegal}, we give a more elaborate example of irreversible dynamics on the dual of the Galilei Lie algebra using a central extension.  

While the finite-dimensional theory provides algebraic guidance for the construction of symmetric brackets, the infinite-dimensional setting is more subtle. Particularly for Poisson manifolds arising from Lie–Poisson structures, there are no general criteria that ensure either the vanishing or the nonvanishing of the second Lichnerowicz–Poisson cohomology group. Unlike the finite-dimensional case, the Whitehead lemmas do not extend to infinite dimensions, and the relationship between Lichnerowicz–Poisson cohomology and Chevalley–Eilenberg cohomology becomes significantly more intricate. This added complexity stems from the use of dual pairings that involve density weights, which influence both the underlying topological structure and the associated cohomology theories. Despite these complications, there is a substantial amount of results in continuous Lie algebra cohomology, particularly for algebras of vector fields over smooth manifolds, that can be used to analyse the Lichnerowicz-Poisson cohomology, such as the monograph \cite{fuks2012cohomology} and research articles by \cite{janssens2015central, janssens2024universal, janssens2025generalized}. In Section \ref{sec:irrevvects1}, we discuss an infinite-dimensional example of irreversible dynamics, in the setting of orientation-preserving diffeomorphisms of the circle using a central extension.

\section{Irreversible dynamics on $\mathfrak{se}(2)^*$}\label{sec:irreversiblese2}
To construct irreversible systems on the dual of a Lie algebra with a Lie-Poisson structure, we require the ability to deform the Lie-Poisson bracket. For the first two example, we take the Lie group $SE(2)$, widely used in planar robotics and control, whose Lie algebra $\mathfrak{se}(2)$ is not semisimple. In the first of these examples, we use a central extension and in the second example we use a generic Poisson 2-cocycle. We first discuss the notation and preliminary definitions.

The Lie group $SE(2) = SO(2) \ltimes \mathbb{R}^2$ is the semidirect product of the abelian rotation group $SO(2)$ and the abelian group of translations $\mathbb{R}^2$. It is defined as the set
\begin{equation}
    SE(2) = \{ (R, \mathbf{v}) \mid R \in SO(2), \mathbf{v} \in \mathbb{R}^2 \},
\end{equation}
equipped with the group multiplication
\begin{equation}
    (R, \mathbf{v}) \cdot (R', \mathbf{v}') = (R R', R \mathbf{v}' + \mathbf{v}).
\end{equation}
It is straightforward to verify that $SE(2)$ is a Lie group under this operation. It is not semisimple, so the Whitehead lemmas do not eliminate the possibility for setting up metriplectic dynamics from a Poisson structure. The group $SE(2)$ admits a faithful matrix representation as a subgroup of $GL(3, \mathbb{R})$
\begin{equation}
    \begin{bmatrix}
        R & \mathbf{v} \\
        \mathbf{0}^T & 1
    \end{bmatrix}
    \begin{bmatrix}
        R' & \mathbf{v}' \\
        \mathbf{0}^T & 1
    \end{bmatrix}
    =
    \begin{bmatrix}
        R R' & R \mathbf{v}' + \mathbf{v} \\
        \mathbf{0}^T & 1
    \end{bmatrix},
\end{equation}
where $R, R' \in SO(2)$, $\mathbf{v}, \mathbf{v}' \in \mathbb{R}^2$, and $\mathbf{0}^T = [0 \quad 0]$ is the zero row vector. This shows that $SE(2)$ is isomorphic to a closed subgroup of $GL(3, \mathbb{R})$. Every rotation matrix $R \in SO(2)$ can be parametrized by a single angle $\varphi \in [0, 2\pi)$ as
\begin{equation}
    R(\varphi) = \begin{bmatrix}
        \cos\varphi & -\sin\varphi \\
        \sin\varphi & \cos\varphi
    \end{bmatrix},
\end{equation}
which means that there is a group isomorphism (covering map) from $SO(2)$ to the unit circle $S^1 \subset \mathbb{C}$, given by
\begin{equation}
    R(\varphi) \mapsto e^{i\varphi}.
\end{equation}
Under this identification, the group multiplication $R' \mapsto R R'$ corresponds to the addition of angles modulo $2\pi$
\begin{equation}
    \varphi' \mapsto \varphi + \varphi' \quad (\text{mod } 2\pi),
\end{equation}
which reveals the abelian nature of $SO(2)$. We also see that the dimension of $SE(2)$ is 3, since $SO(2)$ is 1-dimensional and $\mathbb{R}^2$ is 2-dimensional. The Lie algebra $\mathfrak{se}(2)$ is the Lie algebra of the Lie group $SE(2)$, and consists of matrices of the form
\begin{equation}
    \mathfrak{se}(2) = \left\{
    \begin{bmatrix}
        \omega J & \mathbf{u} \\
        0^T & 0
    \end{bmatrix}
    \,\middle|\,
    \omega \in \mathbb{R},\, \mathbf{u} \in \mathbb{R}^2
    \right\},
\end{equation}
where
\begin{equation}
    J = \begin{bmatrix}
        0 & -1 \\
        1 & 0
    \end{bmatrix}
\end{equation}
is the standard generator of $\mathfrak{so}(2)$. Here, one sees that the space of 2-by-2 traceless skew-symmetric matrices is 1-dimensional. It follows that a convenient basis for $\mathfrak{se}(2)$ is
\begin{equation}\label{eq:se2basis}
    \xi_1 =
    \begin{bmatrix}
        0 & -1 & 0 \\
        1 & 0 & 0 \\
        0 & 0 & 0
    \end{bmatrix}, \quad
    \xi_2 =
    \begin{bmatrix}
        0 & 0 & 1 \\
        0 & 0 & 0 \\
        0 & 0 & 0
    \end{bmatrix}, \quad
    \xi_3 =
    \begin{bmatrix}
        0 & 0 & 0 \\
        0 & 0 & 1 \\
        0 & 0 & 0
    \end{bmatrix}.
\end{equation}
These correspond to infinitesimal rotation, translation in the $x$-direction, and translation in the $y$-direction, respectively.  The Lie brackets in this basis are
\begin{equation}
\begin{aligned}
    [\xi_1, \xi_2]_{\mathfrak{se}(2)} &= \xi_3, \\
    [\xi_1, \xi_3]_{\mathfrak{se}(2)} &= -\xi_2, \\
    [\xi_2, \xi_3]_{\mathfrak{se}(2)} &= 0.
\end{aligned}
\end{equation}
This defines the Lie algebra structure of \( \mathfrak{se}(2) \), reflecting the semidirect product nature: the rotation generator \( \xi_1 \) acts on the translation generators \( \xi_2, \xi_3 \) as the standard action of \( \mathfrak{so}(2) \) on \( \mathbb{R}^2 \). The exponential map \( \exp: \mathfrak{se}(2) \to SE(2) \) maps elements of the Lie algebra to the Lie group. For an element
\begin{equation}
    \xi =
    \begin{bmatrix}
        \omega J & \mathbf{u} \\
        0^T & 0
    \end{bmatrix}
    \in \mathfrak{se}(2),
\end{equation}
the exponential map on $\mathfrak{se}(2)$ admits a closed form expression and is given by
\begin{equation}
    \exp(\xi) =
    \begin{bmatrix}
        \exp(\omega J) & V(\omega) \mathbf{u} \\
        0^T & 1
    \end{bmatrix},
\end{equation}
where \( \exp(\omega J) \in SO(2) \) is the usual matrix exponential and
\begin{equation}
    V(\omega) = \frac{\sin\omega}{\omega}I + \frac{1-\cos\omega}{\omega}J,
\end{equation}
for $\omega\neq 0$, $I$ the 2-by-2 identity matrix and $J$ as before. If $\omega=0$, then $V(\omega)=I$. This formula can be derived from the definition of the exponential as a power series using the Taylor series expansion and the Baker–Campbell–Hausdorff formula. The adjoint action \( \operatorname{Ad}_g : \mathfrak{se}(2) \to \mathfrak{se}(2) \) of the group element \( g = (R, v) \in SE(2) \) on the Lie algebra is defined by
\begin{equation}
    \operatorname{Ad}_g (\xi) = g \xi g^{-1},
\end{equation}
where both \( g \) and \( \xi \) are viewed as 3-by-3 matrices and the product is standard matrix multiplication. Writing \( \xi \in \mathfrak{se}(2) \) in block form,
\begin{equation}
    \xi = 
    \begin{bmatrix}
        \omega J & \mathbf{u} \\
        \mathbf{0}^T & 0
    \end{bmatrix}, \quad
    g = 
    \begin{bmatrix}
        R & \mathbf{v} \\
        \mathbf{0}^T & 1
    \end{bmatrix},
\end{equation}
we compute
\begin{equation}
    \operatorname{Ad}_g (\xi) =
    \begin{bmatrix}
        R \omega J R^{-1} & R \mathbf{u} - R \omega J R^{-1} \mathbf{v} \\
        \mathbf{0}^T & 0
    \end{bmatrix}.
\end{equation}
Using $R^{-1} = R^T$ and $R J R^T = J$, since $SO(2)$ commutes with $J$, we obtain
\begin{equation}
    \operatorname{Ad}_g (\xi) =
    \begin{bmatrix}
        \omega J & R \mathbf{u} - \omega J \mathbf{v} \\
        \mathbf{0}^T & 0
    \end{bmatrix}.
\end{equation}
To understand the geometry of Lie–Poisson dynamics, we study the coadjoint action, which governs how momenta transform under group symmetries. The coadjoint representation $\operatorname{Ad}^*_{g^{-1}} : \mathfrak{se}(2)^* \to \mathfrak{se}(2)^*$ is the algebraic dual of the adjoint representation under the Frobenius pairing. Note that in order for the coadjoint representation to be a left representation, the inverse on the group element is necessary. We will frequently identify elements $\mu \in \mathfrak{se}(2)^*$ with pairs $\mu = (\zeta, \mathbf{p}) \in \mathbb{R} \times \mathbb{R}^2$, where $\zeta \in \mathbb{R}$ is the angular momentum and $p \in \mathbb{R}^2$ is the linear momentum. One can also formulate a matrix representation similar to the adjoint representations. Then the coadjoint action of $ g = (R, v) \in SE(2)$ on $\mu$ is given by
\begin{equation}
    \operatorname{Ad}^*_{g^{-1}} (\mu) =
    \begin{bmatrix} \zeta -\mathbf{v}\cdot (JR\mathbf{p}) &  R \mathbf{p}\\
    \mathbf{0}^T & 1\end{bmatrix}.
\end{equation}
In other words, $\operatorname{Ad}^*_{g^{-1}}(\zeta,\mathbf{p}) = (\zeta-\mathbf{v}\cdot(JR\mathbf{p})+R\mathbf{p}$, from which it is clear that the angular momentum variable $\zeta$ is invariant under the coadjoint action, while the linear momentum $\mathbf{p}$ undergoes a translation and rotation. 

The coadjoint orbit $\mathcal{O}_\mu \subset \mathfrak{se}(2)^*$ through a point $\mu = (\zeta,\mathbf{p})$ is the image of $\mu$ under all coadjoint transformations
\begin{equation}
\mathcal{O}_{(\zeta, \mathbf{p})} = \left\{ \big(\zeta - \mathbf{v} \cdot (J R \mathbf{p}), R \mathbf{p}\big) \mid R \in SO(2), \mathbf{v} \in \mathbb{R}^2 \right\}.
\end{equation}
These coadjoint orbits uniquely characterize the symplectic leaves of the Lie–Poisson structure on $\mathfrak{se}(2)^*$ and are classified as follows
\begin{itemize}
\item If $\mathbf{p} = \mathbf{0}$, then $R\mathbf{p} = \mathbf{0}$, and the translation term vanishes. The orbit is simply $\mathcal{O}{(\zeta,\mathbf{0})} = { (\zeta, \mathbf{0}) }$. These are 0-dimensional points located along the $\zeta$-axis.
\item If $\mathbf{p} \neq \mathbf{0}$, as $R$ varies, $R\mathbf{p}$ traces out a circle of constant radius $\|\mathbf{p}\| > 0$. For any fixed rotation $R$, as the translation $\mathbf{v}$ varies over $\mathbb{R}^2$, the inner product $\mathbf{v} \cdot (J R \mathbf{p})$ independently sweeps out all of $\mathbb{R}$. Thus, the orbit is diffeomorphic to $S^1 \times \mathbb{R}$, which has the structure of a 2-dimensional cylinder.
\end{itemize}
The coadjoint orbits of $SE(2)$ are therefore either points or cylinders. The regular symplectic leaves are the cylinders (where the Casimir $\|\mathbf{p}\|^2 > 0$), while the point orbits residing on the $\zeta$-axis (where $\|\mathbf{p}\|^2 = 0$) constitute the singular leaves of the manifold.

Given an initial condition, Hamiltonian dynamics takes place on the coadjoint orbit containing the initial datum. Hamiltonian dynamics is described by the Lie-Poisson bracket. The Lie--Poisson bracket on $\mathfrak{se}(2)^*$ is defined for smooth functions $f,g\in C^\infty(\mathfrak{se}(2)^*)$ as
\begin{equation}
    \{f, g\}(\mu) = \left\langle \mu, \left[ \frac{\partial f}{\partial \mu}, \frac{\partial g}{\partial \mu} \right]_{\mathfrak{se}(2)} \right\rangle,
\end{equation}
where $\mu \in {\mathfrak{se}(2)}^*$, $\frac{\partial f}{\partial \mu}, \frac{\partial g}{\partial \mu} \in \mathfrak{se}(2)$, $\langle\,,\,\rangle$ is the duality pairing on $\mathfrak{se}(2)^*\times\mathfrak{se}(2)$, and $[\,\cdot\,, \,\cdot\,]_{\mathfrak{se}(2)}$ is the Lie bracket on $\mathfrak{se}(2)$. Using the basis $\{\xi_1, \xi_2, \xi_3\}$ of $\mathfrak{se}(2) $ given in \eqref{eq:se2basis}, with dual coordinates $(\zeta, p_1, p_2)$, the nonzero Lie brackets lead to the following Lie--Poisson bracket structure
\begin{equation}\label{eq:se2pb}
\begin{aligned}
    \{\zeta, p_1\} &= -p_2, \\
    \{\zeta, p_2\} &= p_1, \\
    \{p_1, p_2\} &= 0.
\end{aligned}
\end{equation}
So, in general, for $f, g \in C^\infty(\mathfrak{se}(2)^*)$, the Lie--Poisson bracket is given by
\begin{equation}
    \{f, g\}(\zeta, p_1, p_2) = 
    -p_2 \left( \frac{\partial f}{\partial \zeta} \frac{\partial g}{\partial p_1} - \frac{\partial g}{\partial \zeta} \frac{\partial f}{\partial p_1} \right)
    + p_1 \left( \frac{\partial f}{\partial \zeta} \frac{\partial g}{\partial p_2} - \frac{\partial g}{\partial \zeta} \frac{\partial f}{\partial p_2} \right).
\end{equation}
This bracket is linear in the momenta $(p_1, p_2)$, and defines a Lie–Poisson structure on $\mathfrak{se}(2)^*$. The regular coadjoint orbits are 2-dimensional, meaning that the Poisson bivector associated with the Lie-Poisson bracket has rank 2. This implies that the Poisson tensor has a 1-dimensional kernel almost everywhere, so there exists exactly one functionally independent Casimir, which is given by
\begin{equation}\label{eq:se2casimir}
    C(\zeta, p_1,p_2) = p_1^2 + p_2^2.
\end{equation}
The Casimir is the squared norm of the linear momentum, which defines a $SO(2)$-invariant Euclidean metric on the linear momentum subspace of $\mathfrak{se}(2)^*$. It can therefore be viewed as a degenerate Riemannian structure. The functional independence means that any function of the form $\Phi(p_1^2+p_2^2)$ for a smooth function $\Phi$ is a valid Casimir. To see that this is indeed a Casimir, one checks that $\{C, \zeta\} = \{C, p_1\} = \{C, p_2\} = 0 $, which reflects the fact that the level sets of $p_1^2+p_2^2$ are invariant under the coadjoint action. Given a Hamiltonian $H(\zeta, p_1, p_2)$, the Lie–Poisson equations of motion are:
\begin{equation}
\begin{aligned}
    \frac{\boldsymbol{d}}{\boldsymbol{d}t}{\zeta} &= \{\zeta, H\} = -p_2 \frac{\partial H}{\partial p_1} + p_1 \frac{\partial H}{\partial p_2}, \\
    \frac{\boldsymbol{d}}{\boldsymbol{d}t}{p}_1 &= \{p_1, H\} = p_2 \frac{\partial H}{\partial \zeta}, \\
    \frac{\boldsymbol{d}}{\boldsymbol{d}t}{p}_2 &= \{p_2, H\} = -p_1 \frac{\partial H}{\partial \zeta}.
\end{aligned}
\end{equation}
These equations describe the evolution of momentum and angular momentum in a planar rigid body or control system whose Hamiltonian is invariant under the left action of $SE(2)$. To move towards irreversible dynamics, we declare $S(\zeta,p)=\frac{1}{2}(p_1^2+p_2^2)$ to be the entropy, as this is the only functionally independent Casimir. 

\paragraph{Central extension.}
The Lie algebra $\mathfrak{se}(2)$ is not semisimple and has second cohomology $H^2_{CE}(\mathfrak{se}(2),\mathbb{R})\simeq\mathbb{R}$ (this is explicitly computed in Appendix I in \cite{souriau1970structure}). Given the basis of $\mathfrak{se}(2)$ as in \eqref{eq:se2basis}, a nontrivial central extension is given by including a central element $z$ such that
\begin{equation}
    \begin{aligned}
        [\xi_1,\xi_2] &= \xi_3,\\
        [\xi_1,\xi_3] &= -\xi_2,\\
        [\xi_2,\xi_3] &= z,\\
        [z,-] &= 0.
    \end{aligned}
 \end{equation}
It can be checked that this bracket satisfies the properties to be a Lie bracket. This yields the centrally extended Lie algebra $\widehat{\mathfrak{se}}(2)\simeq \mathbb{R}^4$. A Lie-Poisson structure on the dual of a centrally extended Lie algebra is called an affine Lie-Poisson bracket. The resulting affine Lie-Poisson bracket on $\widehat{\mathfrak{se}}(2)^*\simeq\mathbb{R}^4$ with coordinates $(\zeta,p_1,p_2,c)$ is given by
\begin{equation}
    \begin{aligned}
        \{\zeta,p_1\} &= -p_2,\\
        \{\zeta,p_2\} &= p_1,\\
        \{p_1,p_2\} &= c,
    \end{aligned}
\end{equation}
and all other brackets vanish. In particular, all brackets involving $c$ vanish. This means that $c$ takes on the role of a Casimir of the affine Lie-Poisson bracket. So, in general, for $f,g\in C^\infty(\widehat{\mathfrak{se}}(2)^*)$, the affine Lie-Poisson bracket is given by
\begin{equation}\label{eq:affinelp}
    \{f,g\}(\zeta,p_1,p_2,c) = -p_2\left(\frac{\partial f}{\partial\zeta}\frac{\partial g}{\partial p_1}-\frac{\partial g}{\partial \zeta}\frac{\partial f}{\partial p_1}\right) + p_1\left(\frac{\partial f}{\partial\zeta}\frac{\partial g}{\partial p_2}-\frac{\partial g}{\partial \zeta}\frac{\partial f}{\partial p_2}\right) + c\left(\frac{\partial f}{\partial p_1}\frac{\partial g}{\partial p_2}-\frac{\partial g}{\partial p_1}\frac{\partial f}{\partial p_2}\right)
\end{equation}
The affine Lie-Poisson bracket satisfies the Jacobi identity since the cocycle term satisfies the 2-cocycle identity. This means also that the cocycle term in isolation induces itself a Poisson bracket. The entropy $S(\zeta,p_1,p_2,c) = \frac{1}{2}(p_1^2+p_2^2)$ is not a Casimir of the affine Lie-Poisson bracket, as shown by the following computation. Let $f\in C^\infty(\widehat{\mathfrak{se}}(2)^*)$, then
\begin{equation}\label{eq:centralextlp}
    \begin{aligned}
        \{f,S\} &= \frac{\partial f}{\partial \zeta}(p_1p_2 - p_1p_2) + c\left(p_2\frac{\partial f}{\partial p_1} - p_1\frac{\partial f}{\partial p_2}\right) = c\left(p_2\frac{\partial f}{\partial p_1} - p_1\frac{\partial f}{\partial p_2}\right)
    \end{aligned}
\end{equation}
which does not vanish in general. So one sees that the central extension has changed the kernel of the Poisson bracket. Since $c$ is the variable associated with the central extension, it is constant under evolution by affine Lie-Poisson dynamics. That means that the initial datum specifies $c$ once and for all, and it also implies that $C(\zeta,p_1,p_2,c)=c$ is a Casimir for the affine Lie-Poisson bracket. This is in fact the only Casimir for the affine Lie-Poisson bracket on $\widehat{\mathfrak{se}}(2)^*$, it can be checked by a direct computation that a Casimir must not depend on any of the other coordinates.

Irreversible dynamics can now be constructed by using the Poisson bivectors associated with the Lie-Poisson structure on $\mathfrak{se}(2)^*$ and the affine Lie-Poisson structure on $\widehat{\mathfrak{se}}(2)^*$. Let us denote the Poisson bivector associated with the Lie-Poisson bracket on $\mathfrak{se}(2)^*$ by $\pi$ and the Poisson bivector associated with the affine Lie-Poisson bracket on $\widehat{\mathfrak{se}}(2)^*$ by $\pi_c$. With a slight abuse of notation, the coefficient matrices of the Poisson bivectors are explicitly given by
\begin{equation}
    \pi = \begin{bmatrix}
        0 & -p_2 & p_1 & 0\\
        p_2 & 0 & 0 & 0\\
        -p_1 & 0 & 0 &0\\
        0 & 0 & 0 & 0
    \end{bmatrix}, \qquad \pi_c = \begin{bmatrix}
        0 & -p_2 & p_1 & 0\\
        p_2 & 0 & c & 0\\
        -p_1 & -c & 0 & 0\\
        0 & 0 & 0 & 0
    \end{bmatrix}
\end{equation}
where one may note that $\pi$ can be embedded into $\pi_c$. In other words, given a Hamiltonian $H:\mathfrak{se}(2)^*\to\mathbb{R}$ and the same Hamiltonian extended to $\widehat{\mathfrak{se}}(2)^*$, we have the original Lie-Poisson dynamics on $\mathfrak{se}(2)^*$, affine Lie-Poisson dynamics on $\mathfrak{se}(2)^*$, the original Lie-Poisson dynamics on $\widehat{\mathfrak{se}}(2)^*$ and the affine Lie-Poisson dynamics on $\widehat{\mathfrak{se}}(2)^*$. The subtle difference between these dynamical systems is whether one thinks of the variable associated with the central extension as a parameter or a coordinate, but the dynamics itself is not affected. 

One may also split $\pi_c$ into the enlarged $\pi$ and the cocycle-term that we call $\Theta$, as defined in the following equation
\begin{equation}
    \pi_c = \pi + \Theta = \begin{bmatrix}
        0 & -p_2 & p_1 & 0\\
        p_2 & 0 & 0 & 0\\
        -p_1 & 0 & 0 &0\\
        0 & 0 & 0 & 0
    \end{bmatrix} + \begin{bmatrix}
        0 & 0 & 0 & 0\\
        0 & 0 & c & 0\\
        0 & -c & 0 &0\\
        0 & 0 & 0 & 0
    \end{bmatrix}.
\end{equation} 
One can now also study the dynamics induced by the affine Lie-Poisson bracket, and is discussed in detail in \cite{barbaresco2020lie}. The original Lie-Poisson dynamics on $\mathfrak{se}(2)^*$ is obtained by embedding into $\widehat{\mathfrak{se}}(2)^*$ and taking initial data with $c=0$. The affine Lie-Poisson dynamics on $\widehat{\mathfrak{se}}(2)^*$ is given by
\begin{equation}
    \frac{\boldsymbol{d}}{\boldsymbol{d}t}\left(\begin{matrix}
        \zeta\\
        p_1\\
        p_2\\
        c
    \end{matrix}\right) = \begin{bmatrix}
        0 & -p_2 & p_1 & 0\\
        p_2 & 0 & c & 0\\
        -p_1 & -c & 0 & 0\\
        0 & 0 & 0 & 0
    \end{bmatrix}\left(\begin{matrix}
        \partial H/\partial\zeta\\
        \partial H/\partial p_1\\
        \partial H/\partial p_2\\
        \partial H/\partial c
    \end{matrix}\right)
\end{equation}
It is easy to see that the quantity $c$ does not evolve under the dynamics (this follows immediately from the centrality of $c$), so the initial conditions determine whether one deals with embedded Lie-Poisson dynamics or affine Lie-Poisson dynamics. The symmetric bracket for irreversible dynamics is now obtained from the tensor product $\Theta\otimes\Theta$. 

Given a generator $\mathcal{F}=H+S$, with $H:\widehat{\mathfrak{se}}(2)^*\to\mathbb{R}$ (the original Hamiltonian on $\mathfrak{se}(2)^*$ trivially extended to $\widehat{\mathfrak{se}}(2)^*$ and $S:\widehat{\mathfrak{se}}(2)^*\to\mathbb{R}$ (given by $S=\Phi(p_1^2+p_2^2)$ for some arbitrary smooth function $\Phi$, since it is the only valid entropy in this case), we obtain the following irreversible dynamics
\begin{equation}
    \begin{aligned}
    \frac{\boldsymbol{d}}{\boldsymbol{d}t} f  &= \{H,f\} + ((S,f))\\
    &= -p_2\left(\frac{\partial H}{\partial \zeta}\frac{\partial f}{\partial p_1} - \frac{\partial f}{\partial \zeta}\frac{\partial H}{\partial p_1}\right)+p_1\left(\frac{\partial H}{\partial\zeta}\frac{\partial f}{\partial p_2} - \frac{\partial f}{\partial\zeta}\frac{\partial H}{\partial p_2}\right)\\
    &\qquad + c^2\left(\Big(\frac{\partial S}{\partial p_1}\frac{\partial H}{\partial p_2}-\frac{\partial H}{\partial p_1}\frac{\partial S}{\partial p_2}\Big)\Big(\frac{\partial f}{\partial p_1}\frac{\partial H}{\partial p_2}-\frac{\partial H}{\partial p_1}\frac{\partial f}{\partial p_2}\Big)\right)
    \end{aligned}
\end{equation}
By construction, the brackets satisfy the axioms of irreversible mechanics. Nevertheless, as a final check and a summary, let us establish this explicitly. The entropy is a Casimir of the Lie-Poisson bracket, as established below \eqref{eq:se2casimir}. The entropy is not in the kernel of the symmetric bracket since it is not in the kernel of the centrally extended bracket as shown in \eqref{eq:centralextlp}. Taking $f=S$, we see that $\{H,S\} = 0$ and the evolution of $S$ is given by 
\begin{equation}
\frac{\boldsymbol{d}}{\boldsymbol{d}t} S = c^2(\Phi')^2\left[p_1\frac{\partial H}{\partial p_2} - p_2\frac{\partial H}{\partial p_1}\right]^2 \geq 0.
\end{equation} 
Taking $f=H$, we see that $\{H,H\}=0$, $((S,H)) = 0$, and it follows that $H$ is conserved. For the well-posedness of solutions to the irreversible dynamics, one can use the general theory of ordinary differential equations to obtain local existence and uniqueness of solutions.  To extend these solutions globally, it is mathematically convenient to leverage the monotonic evolution of the entropy. By ensuring the function $-S$ is bounded from below, it can serve as a natural Lyapunov function. The effectiveness of these global continuation techniques depends, of course, on the specific choice of the Hamiltonian and the entropy profile $\Phi$.

To make this example more concrete, we briefly investigate the irreversible version of Kirchhoff's equations for a 2D rigid body moving in an ideal fluid. For a detailed discussion of these equations and their geometric mechanic interpretation, we refer to \cite{vankerschaver2010dynamics}. The Hamiltonian represents the total kinetic energy of the body, including the added mass of the fluid it displaces and is given by
\begin{equation}\label{eq:kirchhoffham}
H(\zeta, p_1, p_2) = \frac{\zeta^2}{2I} + \frac{p_1^2}{2m_1} + \frac{p_2^2}{2m_2},
\end{equation}
where $I$ is the moment of inertia, and $m_1$ and $m_2$ are the effective masses along the principal axes of the body. For the entropy, we take $S(\zeta,p_1,p_2) = \frac{1}{2}(p_1^2+p_2^2)$. We denote the mass anisotropy as $\Delta m = \left(\frac{1}{m_2} - \frac{1}{m_1}\right)$. The irreversible equations of motion are then given by
\begin{equation}
\begin{aligned}
\frac{\boldsymbol{d}\zeta}{\boldsymbol{d}t} &= -\Delta m \, p_1 p_2, \\
\frac{\boldsymbol{d}p_1}{\boldsymbol{d}t} &= -\frac{\zeta}{I} p_2 + c^2 \frac{\Delta m}{m_2} p_1 p_2^2, \\
\frac{\boldsymbol{d}p_2}{\boldsymbol{d}t} &= \frac{\zeta}{I} p_1 - c^2 \frac{\Delta m}{m_1} p_1^2 p_2.
\end{aligned}
\end{equation}
Observe that the dissipation is geometrically locked to the asymmetry of the Hamiltonian. A perfectly circular body moving in the fluid will experience no dissipation from this 2-cocycle, but an asymmetric body will couple the linear momenta, triggering the symmetric bracket to monotonically increase the translational kinetic energy while preserving the total energy $H$.

\paragraph{Generic Poisson 2-cocycle.}
As a second example, we consider a Poisson 2-cocycle not arising from a central extension. To construct such a cocycle, we can exploit the fact that $\mathfrak{se}(2)^*$ is a 3-dimensional manifold, allowing us to use the isomorphism between bivectors and vector fields via standard 3D vector calculus. We can represent any bivector $\varpi$ in the coordinate basis $(\zeta, p_1, p_2)$ as
\begin{equation}
\varpi = \varpi_1\frac{\partial}{\partial \zeta}\wedge \frac{\partial}{\partial p_1} + \varpi_2\frac{\partial}{\partial \zeta}\wedge \frac{\partial}{\partial p_2} + \varpi_3\frac{\partial}{\partial p_1}\wedge\frac{\partial}{\partial p_2}.
\end{equation}
Under the standard isomorphism, this bivector corresponds to a proxy vector field $\varpi=(\varpi_3,-\varpi_2,\varpi_1)^T$. Similarly, the original Lie-Poisson bivector $\pi$ corresponds to a vector field $\pi$. The advantage of this representation is that the Schouten-Nijenhuis bracket translates directly into standard vector operations. Specifically, the condition for $\varpi$ to be a 2-cocycle, $[\pi, \varpi] = 0$, becomes the vector equation
\begin{equation}
	\pi \cdot (\nabla \times \varpi) + \varpi \cdot (\nabla \times \pi) = 0.
\end{equation}
The original Lie-Poisson bracket on $\mathfrak{se}(2)^*$ is defined by the relations \eqref{eq:se2pb}. This yields the proxy vector field $\pi = (0, -p_1, -p_2)^T$. The curl of $\pi$ is zero, hence the cocycle condition amounts to
\begin{equation}
	\pi \cdot (\nabla \times \varpi) = 0.
\end{equation}
To satisfy this, we require the curl of $\varpi$ to be orthogonal to $\pi$. A straightforward choice is to take $\nabla\times\varpi=(0,p_2,-p_1)^T$. A vector field $\varpi$ that satisfies this is given by
\begin{equation}
	\varpi = \Big(\frac{1}{2}p_2^2, -\zeta p_1, 0\Big)^T.
\end{equation}
Translating this back to the bivector components, we find $\varpi_3 = \frac{1}{2}p_2^2$, $\varpi_2 = \zeta p_1$, and $\varpi_1 = 0$, resulting in the 2-cocycle
\begin{equation}
\varpi = \zeta p_1\frac{\partial}{\partial \zeta}\wedge \frac{\partial}{\partial p_2} + \frac{1}{2}p_2^2\frac{\partial}{\partial p_1}\wedge\frac{\partial}{\partial p_2}.
\end{equation}
Finally, to verify that this generic cocycle does not generate a Poisson bracket on its own, we check the quadratic condition $[\varpi,\varpi] = 0$. In our 3D proxy representation, this corresponds to the helicity condition $\varpi \cdot (\nabla \times \varpi) = 0$. Evaluating this expression yields
\begin{equation}
\varpi \cdot (\nabla \times \varpi) = \Big(\frac{1}{2}p_2^2, -\zeta p_1, 0\Big)^T \cdot (0, p_2, -p_1)^T = -\zeta p_1 p_2.
\end{equation}
Since this quantity does not globally vanish, $\varpi$ violates the Jacobi identity. It is therefore strictly a generic 2-cocycle, yet it successfully breaks the Casimir foliation of $\pi$ and serves as a valid geometric generator for the irreversible dynamics. We now define the symmetric bracket as
\begin{equation}
((S,f)) = \varpi(dS, dH)\varpi(df, dH)
\end{equation}
and obtain the irreversible dynamics
\begin{equation}
	\begin{aligned}
		\frac{\boldsymbol{d}}{\boldsymbol{d}t} f  &= \{H,f\} + ((S,f))\\
		&= -p_2\left(\frac{\partial H}{\partial \zeta}\frac{\partial f}{\partial p_1} - \frac{\partial f}{\partial \zeta}\frac{\partial H}{\partial p_1}\right)+p_1\left(\frac{\partial H}{\partial\zeta}\frac{\partial f}{\partial p_2} - \frac{\partial f}{\partial\zeta}\frac{\partial H}{\partial p_2}\right)\\
		&\qquad + \Bigg[\zeta p_1\left(\frac{\partial S}{\partial \zeta}\frac{\partial H}{\partial p_2} - \frac{\partial H}{\partial \zeta}\frac{\partial S}{\partial p_2}\right) + \frac{1}{2}p_2^2\left(\frac{\partial S}{\partial p_1}\frac{\partial H}{\partial p_2}-\frac{\partial H}{\partial p_1}\frac{\partial S}{\partial p_2}\right)\Bigg]\\
		&\qquad \quad \times \Bigg[\zeta p_1\left(\frac{\partial f}{\partial \zeta}\frac{\partial H}{\partial p_2} - \frac{\partial H}{\partial \zeta}\frac{\partial f}{\partial p_2}\right) + \frac{1}{2}p_2^2\left(\frac{\partial f}{\partial p_1}\frac{\partial H}{\partial p_2}-\frac{\partial H}{\partial p_1}\frac{\partial f}{\partial p_2}\right)\Bigg].
	\end{aligned}
\end{equation}
In the above equation, the cross denotes ordinary multiplication. It is immediately clear that the Hamiltonian is preserved. A quick computation shows that the evolution of $S=\Phi(p_1^2+p_2^2)$ is given by
\begin{equation}
	\frac{\boldsymbol{d}}{\boldsymbol{d}t} S = 4(\Phi')^2\left[-\zeta p_1 p_2\frac{\partial H}{\partial \zeta} + \frac{1}{2}p_1p_2^2\frac{\partial H}{\partial p_2} - \frac{1}{2}p_2^3\frac{\partial H}{\partial p_1}\right]^2 \geq 0.
\end{equation} 
We now revisit the irreversible formulation of Kirchhoff's equations for a 2D rigid body moving in an ideal fluid. Given the Hamiltonian as in \eqref{eq:kirchhoffham} and the entropy $S=\frac{1}{2}(p_1^2+p_2^2)$, we obtain for the generic 2-cocycle the following equations of motion
\begin{equation}
\begin{aligned}
\frac{\boldsymbol{d}\zeta}{\boldsymbol{d}t} &= -\Delta m \, p_1 p_2 + p_1 p_2 \left( \frac{1}{2} \Delta m \, p_2^2 - \frac{\zeta^2}{I} \right) \frac{\zeta p_1 p_2}{m_2}, \\
\frac{\boldsymbol{d}p_1}{\boldsymbol{d}t} &= -\frac{\zeta}{I} p_2 + p_1 p_2 \left( \frac{1}{2} \Delta m \, p_2^2 - \frac{\zeta^2}{I} \right)\frac{p_2^3}{2 m_2}, \\
\frac{\boldsymbol{d}p_2}{\boldsymbol{d}t} &= \frac{\zeta}{I} p_1 - p_1 p_2 \left( \frac{1}{2} \Delta m \, p_2^2 - \frac{\zeta^2}{I} \right) p_1 \left( \frac{\zeta^2}{I} + \frac{p_2^2}{2m_1} \right),
\end{aligned}
\end{equation}
where $\Delta m = \frac{1}{m_2} - \frac{1}{m_1}$ denotes the mass anisotropy as before. Observe that for the generic 2-cocycle, the dissipation no longer vanishes if the body is isotropic, i.e., if $\Delta m = 0$.

In the next section, we discuss irreversible dynamics constructed using the central extension of the Galilei algebra.

\section{Irreversible dynamics on $\mathfrak{sgal}(3)^*$}\label{sec:irreversiblegal}
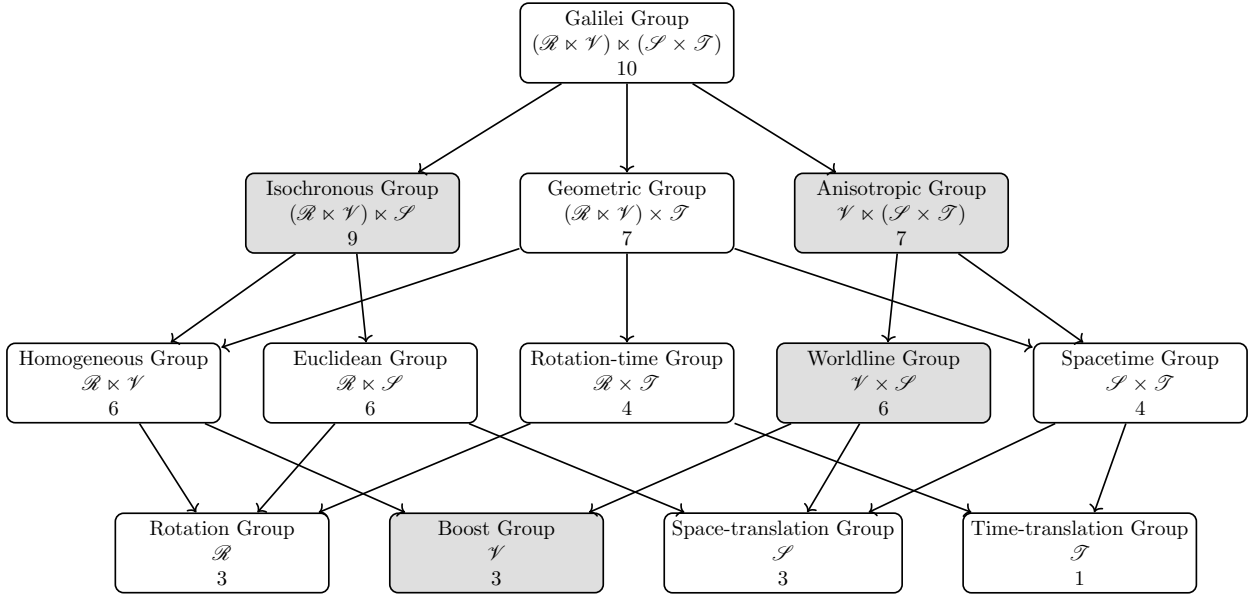
\begin{figure}[H]
\centering
\resizebox{0.95\textwidth}{!}{
\begin{tikzpicture}[
  node distance=1.8cm and 3.0cm,
  shadedbox/.style={draw, rectangle, rounded corners, minimum width=3.6cm, minimum height=1cm, align=center, font=\small, fill=gray!25},
  normalbox/.style={draw, rectangle, rounded corners, minimum width=3.6cm, minimum height=1cm, align=center, font=\small},
  every path/.style={->, thick}
  ]

\node[normalbox] (G) {Galilei Group \\ $(\mathscr{R}\ltimes\mathscr{V})\ltimes(\mathscr{S}\times\mathscr{T})$\\ 10};

\node[shadedbox] (Iso) [below left=1.5cm and 1.0cm of G] {Isochronous Group \\ $(\mathscr{R}\ltimes\mathscr{V})\ltimes\mathscr{S}$ \\ 9};
\node[normalbox] (Geo) [below=1.5cm of G] {Geometric Group \\ $(\mathscr{R}\ltimes\mathscr{V})\times \mathscr{T}$ \\ 7};
\node[shadedbox] (Ani) [below right=1.5cm and 1.0cm of G] {Anisotropic Group \\ $\mathscr{V}\ltimes(\mathscr{S}\times\mathscr{T})$ \\ 7};

\node[normalbox] (RT) [below=1.5cm of Geo] {Rotation-time Group \\ $\mathscr{R}\times\mathscr{T}$ \\ 4};
\node[normalbox] (SE) [left=0.7cm of RT] {Euclidean Group \\ $\mathscr{R}\ltimes\mathscr{S}$ \\ 6};
\node[normalbox] (Hom) [left=0.7cm of SE] {Homogeneous Group \\ $\mathscr{R}\ltimes\mathscr{V}$ \\ 6};
\node[shadedbox] (Wor) [right=0.7cm of RT] {Worldline Group \\ $\mathscr{V}\times\mathscr{S}$ \\ 6};
\node[normalbox] (ST)  [right=0.7cm of Wor] {Spacetime Group \\ $\mathscr{S}\times\mathscr{T}$ \\ 4};

\node[normalbox] (R) [below right=1.5cm and -1.8cm of Hom] {Rotation Group \\ $\mathscr{R}$ \\ 3};
\node[shadedbox] (V) [right=1cm of R] {Boost Group \\ $\mathscr{V}$ \\ 3};
\node[normalbox] (S) [right=1cm of V] {Space-translation Group \\ $\mathscr{S}$ \\ 3};
\node[normalbox] (T) [right=1cm of S] {Time-translation Group \\ $\mathscr{T}$ \\ 1};

\draw (G) -- (Ani);
\draw (G) -- (Geo);
\draw (G) -- (Iso);

\draw (Ani) -- (ST);
\draw (Ani) -- (Wor);
\draw (Geo) -- (ST);
\draw (Geo) -- (RT);
\draw (Geo) -- (Hom);
\draw (Iso) -- (SE);
\draw (Iso) -- (Hom);

\draw (ST) -- (T);
\draw (ST) -- (S);
\draw (Wor) -- (S);
\draw (Wor) -- (V);
\draw (RT) -- (T);
\draw (RT) -- (R);
\draw (SE) -- (S);
\draw (SE) -- (R);
\draw (Hom) -- (V);
\draw (Hom) -- (R);
\end{tikzpicture}
}
\caption{The lattice of subgroups of the Galilei group. To obtain the special Galilei group with its natural action on Euclidean spacetime, one sets $\mathscr{R}=SO(3)$, $\mathscr{V}=\mathbb{R}^3$, $\mathscr{S} = \mathbb{R}^3$, and $\mathscr{T}=\mathbb{R}$. The numbers in the bottom of each box indicate the dimension of the group. The shaded boxes indicate groups that have no relativistic analogue.}
\label{fig:galsubdiag}
\end{figure}

The Galilei group $SGal(3)$ describes the transformations from one inertial frame to another. In an inertial frame, an object with zero net forces acting on it is perceived to move at a constant velocity. Furthermore, the laws of mechanics have the same form in all inertial frames. The action of the Galilei group on space-time $\mathbb{R}\times\mathbb{R}^3$ is given by
\begin{equation}\label{eq:galspacetime}
    (t,\mathbf{x})\mapsto (t+e,A\mathbf{x} + \mathbf{b}t + \mathbf{c})
\end{equation}
where $\mathbf{x}$ denotes displacement, $t$ denotes time, $A\in SO(3)$ is rotation, $\mathbf{b}\in \mathbb{R}^3$ is a boost, $\mathbf{c}\in\mathbb{R}^3$ is a translation in space and $e\in\mathbb{R}$ is a translation in time. Note that we take $SO(3)$ for rotations rather than $O(3)$, so that the Galilei group is connected, hence we write $SGal(3)$ instead of $Gal(3)$. The Galilei group $SGal(3)$ can be identified with the quadruple $(A,\mathbf{b},\mathbf{c},e)$ and is a 10-dimensional noncompact connected Lie group. A convenient matrix representation of the elements of the Galilei group is the following
\begin{equation}\label{eq:linrepgal}
SGal(3) = \left\{ \left(\begin{matrix}
    A & \mathbf{b} & \mathbf{c}\\
    0_{1\times 3} & 1 & e\\
    0_{1\times 3} & 0 & 1
    \end{matrix}\right)\in GL(5,\mathbb{R})\,\Bigg|\,A\in SO(3), \mathbf{b}\in\mathbb{R}^3, \mathbf{c}\in\mathbb{R}^3, e\in\mathbb{R}\right\}.
\end{equation}
For a concise notation, we write $0_{3\times 1}$ for the column vector of zeros, hence $0_{1\times 3}$ is then a row vector of zeros. The Galilei group has many subgroups that themselves play important roles in geometric mechanics. In \cite{levy1971galilei} (Fig. 1), one can find the diagram illustrated in Figure \ref{fig:galsubdiag} that gives a clean overview of the subgroups of the Galilei group.

The anisotropic group (in the second row on the right in Figure \ref{fig:galsubdiag}) is sometimes referred to as the parabolic subgroup of the Galilei group. One typically writes $\mathbb{R}^4=\mathbb{R}^3\times\mathbb{R}$ to denote spacetime. The Galilei group is a semidirect product Lie group consisting of two nested semidirect products: $SGal(3)= (SO(3)\ltimes\mathbb{R}^3)\ltimes\mathbb{R}^4$. In the matrix representation \eqref{eq:linrepgal}, the group multiplication is matrix multiplication and the inverse of an element in $SGal(3)$ has the form
\begin{equation}
    \left(\begin{matrix}
    A & \mathbf{b} & \mathbf{c}\\
    0_{1\times 3} & 1 & e\\
    0_{1\times 3} & 0 & 1
    \end{matrix}\right)^{-1} = \left(\begin{matrix}
    A^{-1} & -A^{-1}\mathbf{b} & -A^{-1}(\mathbf{c}-\mathbf{b}e)\\
    0_{1\times 3}& 1 & -e\\
    0_{1\times 3} & 0 & 1
    \end{matrix}\right)
\end{equation}
The matrix representation of the Galilean Lie algebra $\mathfrak{sgal}(3)$ is given by matrices of the form
\begin{equation}\label{eq:galalglinrep}
    \mathfrak{sgal}(3) = \left\{ \left(\begin{matrix}
    \Xi & \boldsymbol \beta & \boldsymbol \gamma\\
    0_{1\times 3} & 0 & \varepsilon\\
    0_{1\times 3} & 0 & 0
    \end{matrix}\right)\in \mathfrak{gl}(5,\mathbb{R})\,\Bigg|\,\Xi\in\mathfrak{so}(3), \boldsymbol \beta\in\mathbb{R}^3, \boldsymbol\gamma\in\mathbb{R}^3, \varepsilon\in\mathbb{R}\right\}.
\end{equation}
Here $\mathfrak{gl}(5,\mathbb{R})$ denotes the Lie algebra of the real general linear group in dimension five, which consists of all $5\times 5$ matrices with entries in the real numbers, and $\mathfrak{so}(3)$ is the Lie algebra of $3\times 3$ skew-symmetric matrices. This is a 10-dimensional Lie algebra and its basis is easily obtained from its semidirect product components. Using the matrix representation of the Galilean Lie algebra, we compute the adjoint representation of the group to be
\begin{equation}\label{eq:Adgal}
    {\rm Ad}_{(A,\mathbf{b},\mathbf{c},e)}(\widetilde{\Xi}, \widetilde{\boldsymbol{\beta}},\widetilde{\boldsymbol{\gamma}},\widetilde{\varepsilon}) = \left(\begin{matrix}
        A\widetilde{\Xi} A^{-1} & -A\widetilde{\Xi}A^{-1}\mathbf{b} + A\widetilde{\boldsymbol{\beta}} & -A\widetilde{\Xi}A^{-1}(\mathbf{c}-\mathbf{b}e) + A(\widetilde{\boldsymbol{\gamma}} -\widetilde{\boldsymbol{\beta}}e) + \mathbf{b}\widetilde{\varepsilon} \\
        0_{1\times 3} & 1 & \widetilde{\varepsilon} \\
        0_{1\times 3} & 0 & 1
    \end{matrix}\right).
\end{equation}
The adjoint representation of the Galilei algebra is its Lie bracket, and it has the typical form of a nested semidirect product Lie bracket, given by
\begin{equation}\label{eq:adgal}
    {\rm ad}_{(\Xi,\boldsymbol{\beta},\boldsymbol{\gamma},\varepsilon)}(\widetilde{\Xi}, \widetilde{\boldsymbol{\beta}},\widetilde{\boldsymbol{\gamma}},\widetilde{\varepsilon}) = \left(\begin{matrix}
        \Xi\widetilde{\Xi}-\widetilde{\Xi}\Xi & \Xi\widetilde{\boldsymbol{\beta}} - \widetilde{\Xi}\boldsymbol{\beta} & \Xi\widetilde{\boldsymbol{\gamma}} - \widetilde{\Xi}\boldsymbol{\gamma} + \boldsymbol{\beta}\widetilde{\varepsilon} - \widetilde{\boldsymbol{\beta}}\varepsilon\\
        0_{1\times 3} & 0 & 0\\
        0_{1\times 3} & 0 & 0
    \end{matrix}\right).
\end{equation}
To compute the coadjoint representations of the group and its algebra, we require a duality pairing on the algebra. The natural pairing on matrix Lie algebras is the Frobenius pairing. For square matrices $A,B$, the Frobenius pairing is given by $\langle A, B\rangle = {\rm tr}(A^TB)$. The space $\mathfrak{sgal}(3)^*$, the dual of $\mathfrak{sgal}(3)$, is given by matrices of the form
\begin{equation}
    \mathfrak{gal}(3)^* = \left\{ \left(\begin{matrix}
    Z & \mathbf{g} & \mathbf{p}\\
    0_{1\times 3} & 0 & E\\
    0_{1\times 3} & 0 & 0
    \end{matrix}\right)\in \mathfrak{gl}(5,\mathbb{R})\,\Bigg|\,Z\in\mathfrak{so}(3)^*, \mathbf{g}\in\mathbb{R}^3, \textbf{p}\in\mathbb{R}^3, E\in\mathbb{R}\right\},
\end{equation}
with the interpretation that the skew-symmetric matrix $Z$ represents angular momentum, $\mathbf{g}$ represents center of mass, $\mathbf{p}$ represents linear momentum and $E$ represents energy. Given the structure of the matrix representations of $\mathfrak{sgal}(3)$ and $\mathfrak{sgal}(3)^*$, the Frobenius pairing splits up naturally into the four constituents of the semidirect product of the Galilei group
\begin{equation}\label{eq:frobpair}
    \left\langle\left(\begin{matrix}
        Z & \mathbf{g} & \mathbf{p} \\ 
        0_{1\times 3} & 0 & E\\
        0_{1\times 3} & 0 & 0
    \end{matrix}\right),\left(\begin{matrix}
        \Xi & \boldsymbol{\beta} & \boldsymbol{\gamma}\\
        0_{1\times 3} & 0 & \varepsilon \\
        0_{1\times 3} & 0 & 0
    \end{matrix}\right)\right\rangle = \langle Z,\Xi\rangle + \mathbf{g}\cdot\boldsymbol{\beta} + \mathbf{p}\cdot\boldsymbol{\gamma} + E\varepsilon,
\end{equation}
where $\langle\,\cdot\,,\,\cdot\,\rangle$ is again the Frobenius pairing, but now for skew-symmetric matrix $\Xi$ and its dual $Z$, and the dot product denotes the natural pairing on $\mathbb{R}^3$. This is convenient because it permits us to compute the coadjoint actions term by term. The coadjoint representation of the group is given by
\begin{equation}\label{eq:Ad*gal}
    {\rm Ad}^*_{(A,\mathbf{b},\mathbf{c},e)^{-1}}(Z,\mathbf{g},\mathbf{p},E) = \left(\begin{matrix}
        A Z A^{-1} + {\rm skew}\big(\mathbf{b}\otimes (A\mathbf{g} - A\mathbf{p}e) + \mathbf{c}\otimes A\mathbf{p}\big) & A\mathbf{g} - A\mathbf{p}e & A\mathbf{p} \\
        0_{1\times 3} & 0 & E + \mathbf{b}\cdot\mathbf{p} \\
        0_{1\times 3} & 0 & 0
    \end{matrix}\right),
\end{equation}
where $\mathrm{skew}$ extracts the skew-symmetric part of the matrix it acts on. The coadjoint representation of the algebra is given by
\begin{equation}\label{eq:ad*gal}
    {\rm ad}^*_{(\Xi,\boldsymbol{\beta},\boldsymbol{\gamma},\varepsilon)}(Z,\mathbf{g},\mathbf{p},E) = \left(\begin{matrix}
        Z\Xi-\Xi Z - {\rm skew}\big(\boldsymbol{\beta} \otimes \mathbf{g} + \boldsymbol{\gamma}\otimes \mathbf{p}\big) & -\Xi\mathbf{g} + \mathbf{p}\varepsilon & -\Xi\mathbf{p} \\
        0_{1\times 3} & 0 & -\boldsymbol{\beta}\cdot \mathbf{p} \\
        0_{1\times 3} & 0 & 0
    \end{matrix}\right).
\end{equation}
The adjoint and coadjoint representations \eqref{eq:Adgal}, \eqref{eq:adgal}, \eqref{eq:Ad*gal}, \eqref{eq:ad*gal} in case of $SGal(n)$ and $\mathfrak{sgal}(n)$ are obtained mutatis mutandis. In the special case of $n=3$, there exists a convenient isomorphism between $\mathfrak{so}(3)$ and $\mathbb{R}^3$. This isomorphism is the hat map and it sends $\mathbb{R}^3\to\mathfrak{so}(3), \boldsymbol{\xi}\mapsto\widehat{\boldsymbol{\xi}} = \Xi$. Using the hat map and the identification of $\mathfrak{so}(3)^*\simeq\mathfrak{so}(3)$, we have $Z = \widehat{\boldsymbol{\zeta}}$ and $\Xi = \widehat{\boldsymbol{\xi}}$ with $\boldsymbol{\zeta},\boldsymbol{\xi}\in\mathbb{R}^3$. The coadjoint action of the Galilei group on the dual of its Lie algebra is then given by
\begin{equation}\label{eq:ad*galhat}
    {\rm ad}^*_{(\boldsymbol{\xi},\boldsymbol{\beta},\boldsymbol{\gamma},\varepsilon)}(\boldsymbol{\zeta},\mathbf{g},\mathbf{p},E) = \left(\begin{matrix}
        -\boldsymbol{\xi}\times\boldsymbol{\zeta} - \boldsymbol{\beta}\times\mathbf{g} - \boldsymbol{\gamma}\times\mathbf{p} & -\boldsymbol{\xi}\times\mathbf{g} + \mathbf{p}\varepsilon & -\boldsymbol{\xi}\times\mathbf{p} \\
        0_{1\times 3} & 0 & -\boldsymbol{\beta}\cdot \mathbf{p} \\
        0_{1\times 3} & 0 & 0
    \end{matrix}\right).
\end{equation}
We now have explicit forms of all adjoint and coadjoint representations. From the coadjoint representation of the algebra, one obtains the Lie-Poisson bracket on $\mathfrak{sgal}(3)^*$. As a result of the hat map isomorphism, we have several ways of expressing the Lie-Poisson bracket. The first expression is the most common way of writing a Lie-Poisson bracket. The second expression writes the Poisson bivector in an operator form using $\mathbb{R}^3$ vectors, cross products and dot products. Since only $3\times 1$ vectors appear in this formulation, we write $\mathbf{0}$ for $0_{3\times 1}$, which gives an elegant formulation. We use this formulation to show what the Casimirs of the Lie-Poisson bracket are. The third expression is in terms of skew-symmetric matrices with standard matrix-vector multiplication. Here, we distinguish explicitly between the zero matrix, zero vector, its transpose, and scalar zero for full dimensional correctness.
\begin{equation}\label{eq:gallp}
\begin{aligned}
\{f,g\}(\boldsymbol{\zeta}, \mathbf{g}, \mathbf{p}, E) 
&= \boldsymbol{\zeta} \cdot \left( \frac{\partial f}{\partial \boldsymbol{\zeta}} \times \frac{\partial g}{\partial \boldsymbol{\zeta}} \right) + \mathbf{g} \cdot \left( \frac{\partial f}{\partial \boldsymbol{\zeta}} \times \frac{\partial g}{\partial \mathbf{g}} - \frac{\partial g}{\partial \boldsymbol{\zeta}} \times \frac{\partial f}{\partial \mathbf{g}} \right) \\
&\quad + \mathbf{p} \cdot \left( \frac{\partial f}{\partial \boldsymbol{\zeta}} \times \frac{\partial g}{\partial \mathbf{p}} - \frac{\partial g}{\partial \boldsymbol{\zeta}} \times \frac{\partial f}{\partial \mathbf{p}} \right) + \mathbf{p} \cdot \left( \frac{\partial f}{\partial \mathbf{g}} \frac{\partial g}{\partial E} - \frac{\partial g}{\partial \mathbf{g}} \frac{\partial f}{\partial E} \right),\\[1ex]
&= \begin{pmatrix}
    \frac{\partial f}{\partial \boldsymbol{\zeta}}\\[5pt]
    \frac{\partial f}{\partial \mathbf{g}}\\[5pt]
    \frac{\partial f}{\partial \mathbf{p}}\\[5pt]
    \frac{\partial f}{\partial E}
\end{pmatrix}^T
\begin{pmatrix}
    \boldsymbol{\zeta}\times & \mathbf{g}\times & \mathbf{p}\times & \mathbf{0} \\[5pt]
    -\mathbf{g}\times & \mathbf{0}\times & \mathbf{0}\times & \mathbf{p} \\[5pt]
    -\mathbf{p}\times & \mathbf{0}\times & \mathbf{0}\times & \mathbf{0} \\[5pt]
    \mathbf{0}\cdot & -\mathbf{p}\cdot & \mathbf{0}\cdot & 0
\end{pmatrix}
\begin{pmatrix}
    \frac{\partial g}{\partial \boldsymbol{\zeta}}\\[5pt]
    \frac{\partial g}{\partial \mathbf{g}}\\[5pt]
    \frac{\partial g}{\partial \mathbf{p}}\\[5pt]
    \frac{\partial g}{\partial E}
\end{pmatrix},\\
&= 
\begin{pmatrix}
    \frac{\partial f}{\partial \boldsymbol{\zeta}}\\[5pt]
    \frac{\partial f}{\partial \mathbf{g}}\\[5pt]
    \frac{\partial f}{\partial \mathbf{p}}\\[5pt]
    \frac{\partial f}{\partial E}
\end{pmatrix}^T
\begin{pmatrix}
    \widehat{\boldsymbol{\zeta}} & \widehat{\mathbf{g}} & \widehat{\mathbf{p}} & 0_{3\times 1} \\[5pt]
    -\widehat{\mathbf{g}} & 0_{3\times 3} & 0_{3\times 3} & \mathbf{p} \\[5pt]
    -\widehat{\mathbf{p}} & 0_{3\times 3} & 0_{3\times 3} & 0_{3\times 1} \\[5pt]
    0_{1\times 3} & -\mathbf{p}^T & 0_{1\times 3} & 0
\end{pmatrix}
\begin{pmatrix}
    \frac{\partial g}{\partial \boldsymbol{\zeta}}\\[5pt]
    \frac{\partial g}{\partial \mathbf{g}}\\[5pt]
    \frac{\partial g}{\partial \mathbf{p}}\\[5pt]
    \frac{\partial g}{\partial E}
\end{pmatrix}
\end{aligned}
\end{equation}
The third form is the most suitable for an unambiguous expression of the Poisson bivector that defines the Lie-Poisson bracket.

The next task is to find the Casimirs of the Lie-Poisson bracket \eqref{eq:gallp}. For this we need to find a function $C(\boldsymbol{\zeta},\mathbf{g},\mathbf{p},E):\mathfrak{sgal}(3)^*\to\mathbb{R}$ such that
\begin{equation}
    \begin{aligned}
        \boldsymbol{\zeta}\times\frac{\partial C}{\partial\boldsymbol{\zeta}} + \mathbf{g}\times\frac{\partial C}{\partial \mathbf{g}} + \mathbf{p}\times\frac{\partial C}{\partial \mathbf{p}} &= \mathbf{0},\\
        -\mathbf{g}\times\frac{\partial C}{\partial \boldsymbol{\zeta}}+ \mathbf{p}\frac{\partial C}{\partial E} &= \mathbf{0},\\
        -\mathbf{p}\times\frac{\partial C}{\partial \boldsymbol{\zeta}} &= \mathbf{0},\\
        -\mathbf{p}\cdot \frac{\partial C}{\partial\mathbf{g}} &= 0,
    \end{aligned}
\end{equation}
are all simultaneously satisfied for arbitrary $\boldsymbol{\zeta},\mathbf{g},\mathbf{p},E\in\mathbb{R}^3\times \mathbb{R}^3\times \mathbb{R}^3\times\mathbb{R}$. To find such functions, we take a step by step approach to find restrictions on $C$. We focus on the second and third equations. The third equation implies that $\frac{\partial C}{\partial\boldsymbol{\zeta}}$ is collinear with $\mathbf{p}$, which means that $\frac{\partial C}{\partial\boldsymbol{\zeta}} = \lambda\mathbf{p}$ for some $\lambda\in\mathbb{R}$. Substituting this into the second equation, we obtain that $\lambda(\mathbf{g}\times\mathbf{p})+\frac{\partial C}{\partial E}\mathbf{p}=0$. Taking the cross product of this equation with $\mathbf{p}$ yields $\lambda(\mathbf{p}\times(\mathbf{g}\times\mathbf{p})) = -\lambda(\|\mathbf{p}\|^2\mathbf{g} - (\mathbf{p}\cdot\mathbf{g})\mathbf{p})=0$, which is satisfied for $\lambda\neq 0$ if $\mathbf{g}$ is collinear with $\mathbf{p}$. Since $\mathbf{g}$ and $\mathbf{p}$ are arbitrary, this is in general not the case, so $\lambda=0$ must hold. This implies that $C$ is not a function of $\boldsymbol{\zeta}$, and the second equation then implies that $C$ is also not a function of $E$.

We can limit our search to Casimir functions of the form $C(\mathbf{g},\mathbf{p})$. The defining equations reduce to the system
\begin{equation}
	\begin{aligned}
		\mathbf{g}\times\frac{\partial C}{\partial \mathbf{g}} + \mathbf{p}\times\frac{\partial C}{\partial \mathbf{p}} &= \mathbf{0},\\
		\mathbf{p}\cdot \frac{\partial C}{\partial\mathbf{g}} &= 0.
	\end{aligned}
\end{equation}
The first equation is the infinitesimal generator of diagonal $SO(3)$ rotations. It tells us that $C$ must be invariant under simultaneous spatial rotations of $\mathbf{g}$ and $\mathbf{p}$. By classical invariant theory, see e.g. \cite[Section 5.4.3]{goodman2009symmetry}, any such scalar function must be expressible entirely in terms of the fundamental scalar products $x = \mathbf{g}\cdot\mathbf{g}$, $y = \mathbf{p}\cdot\mathbf{p}$, and $z = \mathbf{p}\cdot\mathbf{g}$. We can then write $C = \tilde{C}(x,y,z)$. By the chain rule, the gradient with respect to $\mathbf{g}$ becomes $\frac{\partial C}{\partial \mathbf{g}} = 2\frac{\partial \tilde{C}}{\partial x}\mathbf{g} + \frac{\partial \tilde{C}}{\partial z}\mathbf{p}$. Substituting this into the final remaining constraint, we obtain a linear partial differential equation in terms of the invariant variables
\begin{equation}
	2z\frac{\partial \tilde{C}}{\partial x} + y\frac{\partial \tilde{C}}{\partial z} = 0.
\end{equation}
We can solve the above equation by the method of characteristics. The characteristic equations for this PDE are given by
\begin{equation}
	\frac{dx}{2z} = \frac{dz}{y}, \qquad \frac{dy}{dt} = 0.
\end{equation}
The second characteristic equation trivially dictates that $y$ is a constant of motion along the characteristics. This yields our first independent Casimir, $C_1(\mathbf{p}) = y = \|\mathbf{p}\|^2$. Treating $y$ as a constant parameter, we integrate the first characteristic equation, $y dx = 2z dz$, to find the second constant of motion 
\begin{equation}
	y x = z^2 + c_2 \quad \implies \quad c_2 = x y - z^2.
\end{equation}
Substituting the original definitions of the invariants back into this second constant of integration yields $c_2 = \|\mathbf{g}\|^2\|\mathbf{p}\|^2 - (\mathbf{p}\cdot\mathbf{g})^2$, which via Lagrange's identity is exactly $C_2(\mathbf{g},\mathbf{p}) = \|\mathbf{p}\times\mathbf{g}\|^2$. The method of characteristics explicitly exhausts the degrees of freedom of the PDE system, so we have shown $C_1$ and $C_2$ constitute the full set of functionally independent Casimirs. Given that the dimension of the special Galilei group is 10, and hence $\mathfrak{sgal}(3)^*$ is 10-dimensional, the existence of exactly two functionally independent Casimirs implies that the generic coadjoint orbits of the special Galilei group have dimension 8.

The interpretation of the Casimirs is the following. $C_1$ is the squared norm of the linear momentum, which can be viewed in physical terms as a kinetic energy and in Riemannian terms as a lower rank metric, similar to the Casimir for the Lie-Poisson bracket of $\mathfrak{se}(2)^*$. $C_2$ is known in geometric optics as the skewness, or as the Petzval invariant, see \cite{wolf2004geometric, holm2008geometric1}. 

For irreversible dynamics on $\mathfrak{sgal}(3)^*$, we now have the general expression for an entropy: we can take $S(\boldsymbol{\zeta},\mathbf{g},\mathbf{p},E) = \Phi_1(\|\mathbf{p}\|^2)+\Phi_2(\|\mathbf{p}\times\mathbf{g}\|^2)$, where $\Phi_1$ and $\Phi_2$ are arbitrary smooth functions. Next, we need to find a Lichnerowicz-Poisson 2-cocycle to deform the Lie-Poisson structure on $\mathfrak{sgal}(3)^*$. In \cite{bargmann1954unitary}, it is shown that the second Chevalley-Eilenberg cohomology group $H^2_{CE}(\mathfrak{sgal}(3))$ is 1-dimensional, so there exists a nontrivial 2-cocycle. A detailed derivation of the cohomology of the Galilei group is also presented in \cite[Chapter 12]{souriau1970structure}. We enlarge the Galilei algebra by introducing a central element $z$ such that the Lie brackets involving the three basis elements $\xi_{\boldsymbol{\beta}}$ associated with $\boldsymbol{\beta}$ and the three basis elements $\xi_{\boldsymbol{\gamma}}$ associated with $\boldsymbol{\gamma}$ in \eqref{eq:galalglinrep} satisfy
\begin{equation}
    \begin{aligned}
    [(\xi_{\boldsymbol{\beta}})_i,(\xi_{\boldsymbol{\gamma}})_j] &= z\delta_{ij},\\
    [z,-] &= 0.
    \end{aligned}
\end{equation}
This central extension defines the Bargmann algebra $\mathfrak{b}(3):=\widehat{\mathfrak{sgal}}(3)=((\mathfrak{so}(3)\ltimes\mathbb{R}^3)\ltimes\mathbb{R}^4)\oplus\mathbb{R}$. The Bargmann algebra is an 11-dimensional Lie algebra consisting of quintuples $(\Xi,\boldsymbol{\beta},\boldsymbol{\gamma},\varepsilon,z)$ with $\Xi\in\mathfrak{so}(3)$, $\boldsymbol{\beta}\in\mathbb{R}^3$, $\boldsymbol{\gamma}\in\mathbb{R}^3$, $\varepsilon\in\mathbb{R}$, and $z\in\mathbb{R}$. Recall that the dual variables to $\boldsymbol{\beta}$ and $\boldsymbol{\gamma}$ are $\mathbf{g}$ and $\mathbf{p}$, respectively. Hence, the Lichnerowicz-Poisson 2-cocycle will involve the centre of mass $\mathbf{g}$ and the linear momentum $\mathbf{p}$. Let us denote the dual variable to the central variable $z$ by $M$. This is a conventional choice, since the physical interpretation of $M$ is the mass. Using the hat map isomorphism to set $\mathfrak{so}(3)\simeq\mathbb{R}^3$, we can write for elements of the dual $\mathfrak{b}(3)^*$ the following quintuples $(\boldsymbol{\zeta},\mathbf{g},\mathbf{p},E,M)$. The affine Lie-Poisson bracket on $\mathfrak{b}(3)^*$ is given by
\begin{equation}\label{eq:bargmannalp}
\begin{aligned}
    \{f,g\}(\boldsymbol{\zeta}, \mathbf{g}, \mathbf{p}, E, M) 
&= \boldsymbol{\zeta} \cdot \left( \frac{\partial f}{\partial \boldsymbol{\zeta}} \times \frac{\partial g}{\partial \boldsymbol{\zeta}} \right) + \mathbf{g} \cdot \left( \frac{\partial f}{\partial \boldsymbol{\zeta}} \times \frac{\partial g}{\partial \mathbf{g}} - \frac{\partial g}{\partial \boldsymbol{\zeta}} \times \frac{\partial f}{\partial \mathbf{g}} \right) \\
&\quad + \mathbf{p} \cdot \left( \frac{\partial f}{\partial \boldsymbol{\zeta}} \times \frac{\partial g}{\partial \mathbf{p}} - \frac{\partial g}{\partial \boldsymbol{\zeta}} \times \frac{\partial f}{\partial \mathbf{p}} \right) + \mathbf{p} \cdot \left( \frac{\partial f}{\partial \mathbf{g}} \frac{\partial g}{\partial E} - \frac{\partial g}{\partial \mathbf{g}} \frac{\partial f}{\partial E} \right)\\
&\quad + M\left(\frac{\partial f}{\partial\mathbf{g}}\cdot\frac{\partial g}{\partial\mathbf{p}} - \frac{\partial g}{\partial\mathbf{g}}\cdot \frac{\partial f}{\partial\mathbf{p}}\right).
\end{aligned}
\end{equation}
In Poisson bivector form, using the hat map to go back to the matrix formulation, we have 
\begin{equation}\label{eq:bargmannmat}
    \{f,g\}(\boldsymbol{\zeta}, \mathbf{g}, \mathbf{p}, E, M)  = \begin{pmatrix}
    \frac{\partial f}{\partial \boldsymbol{\zeta}}\\[5pt]
    \frac{\partial f}{\partial \mathbf{g}}\\[5pt]
    \frac{\partial f}{\partial \mathbf{p}}\\[5pt]
    \frac{\partial f}{\partial E}\\[5pt]
    \frac{\partial f}{\partial M}
\end{pmatrix}^T
\begin{pmatrix}
    \widehat{\boldsymbol{\zeta}} & \widehat{\mathbf{g}} & \widehat{\mathbf{p}} & \mathbf{0} & \mathbf{0} \\[5pt]
    -\widehat{\mathbf{g}} & \mathbf{0} & M\mathbb{I} & \mathbf{p} & \mathbf{0}\\[5pt]
    -\widehat{\mathbf{p}} & -M\mathbb{I} & \mathbf{0} & \mathbf{0} & \mathbf{0}\\[5pt]
    \mathbf{0} & -\mathbf{p}^T & \mathbf{0} & 0 & 0\\[5pt]
    \mathbf{0} & \mathbf{0} & \mathbf{0}& 0 & 0
\end{pmatrix}
\begin{pmatrix}
    \frac{\partial g}{\partial \boldsymbol{\zeta}}\\[5pt]
    \frac{\partial g}{\partial \mathbf{g}}\\[5pt]
    \frac{\partial g}{\partial \mathbf{p}}\\[5pt]
    \frac{\partial g}{\partial E}\\[5pt]
    \frac{\partial g}{\partial M}
\end{pmatrix},
\end{equation}
where $\mathbb{I}$ denotes the $3\times 3$ identity matrix. The equations that need to be satisfied for a function $\widehat{C}(\boldsymbol{\zeta},\mathbf{g},\mathbf{p},E,M)$ are
\begin{equation}
    \begin{aligned}
        \boldsymbol{\zeta}\times\frac{\partial \widehat{C}}{\partial \boldsymbol{\zeta}} + \mathbf{g}\times\frac{\partial\widehat{C}}{\partial \mathbf{g}} + \mathbf{p}\times\frac{\partial \widehat{C}}{\partial \mathbf{p}} &= 0,\\
        -\mathbf{g}\times\frac{\partial \widehat{C}}{\partial\boldsymbol{\zeta}} + M\frac{\partial \widehat{C}}{\partial\mathbf{p}} + \mathbf{p}\frac{\partial \widehat{C}}{\partial E} &= 0,\\
        -\mathbf{p}\times\frac{\partial \widehat{C}}{\partial \boldsymbol{\zeta}} - M\frac{\partial \widehat{C}}{\partial\mathbf{g}} &= 0,\\
        -\mathbf{p}\cdot\frac{\partial \widehat{C}}{\partial\mathbf{g}}&= 0.
    \end{aligned}
\end{equation} 
It is shown in \cite{levy1971galilei} (equations 3.27a-c) and also in \cite{fushchich1980reduction} (equation 2.6) that the Casimirs of the affine Lie-Poisson bracket \eqref{eq:bargmannalp}, \eqref{eq:bargmannmat} are given by
\begin{equation}
    \begin{aligned}
        \widehat{C}_1(\boldsymbol{\zeta}, \mathbf{g}, \mathbf{p}, E, M) &= M,\\
        \widehat{C}_2(\boldsymbol{\zeta}, \mathbf{g}, \mathbf{p}, E, M) &= 2ME-\|\mathbf{p}\|^2,\\
        \widehat{C}_3(\boldsymbol{\zeta}, \mathbf{g}, \mathbf{p}, E, M) &= \|M\boldsymbol{\zeta} -\mathbf{g}\times\mathbf{p}\|^2.
    \end{aligned}
\end{equation}
It can be checked via direct computation that these are indeed Casimirs of the affine Lie-Poisson bracket on $\mathfrak{b}(3)^*$. The computation for $\widehat{C}_1$ is trivial, since the partial derivative with respect to $M$ does not play a role in the dynamics. For $\widehat{C}_2$, the third and fourth equations are trivially satisfied, and the first and second equations can easily be verified to hold. For $\widehat{C}_3$, multiple uses of the triple vector product and some careful bookkeeping show that $\widehat{C}_3$ is also a Casimir. 

The physical interpretation of these Casimirs is incredibly rich and has implications for classical mechanics, statistical mechanics and quantum mechanics. The first Casimir is the mass. The second Casimir the nonrelativistic analogue of the mass-shell invariant, which can viewed as defining kinetic energy as an energy. The third Casimir can be viewed as the squared norm of a nonrelativistic analogue of the Pauli-Lubanski pseudovector that is encountered in relativistic mechanics and its interpretation is to define the relation between angular momentum and linear momentum. For the present purposes, the key observation is that none of these Casimirs coincide with the Casimirs of the Lie-Poisson bracket on $\mathfrak{sgal}(3)^*$. This is quick to see, since for $M=0$, the Casimirs of the affine Lie-Poisson bracket on $\mathfrak{b}(3)^*$ coincide with the Casimirs of the Lie-Poisson bracket on $\mathfrak{sgal}(3)^*$. 

In quantum mechanics, Bargmann's theorem \cite{bargmann1954unitary}, interpreted through the lens of geometric quantisation, asserts a fundamental link between projective and unitary representations of Lie groups. Specifically, it states that every projective unitary representation of a connected and simply connected Lie group arises from a true unitary representation of a suitable central extension of the group. In the present context, Bargmann's theorem ensures that any projective unitary representation of the simply connected Galilei group lifts to a true unitary representation of its central extension, the Bargmann group. This central extension incorporates mass as a central charge, and its existence is essential to consistently implement Galilean symmetry in quantum mechanics. In statistical mechanics, this structure has concrete implications: one obtains Galilean invariance of the partition function only if the central extension is properly taken into account. That is, the mass must be included as a fundamental parameter that determines the transformation properties of states and ensures the correct symmetry behavior of thermodynamic quantities.

We now have verified all the necessary conditions for irreversible dynamics on the dual of the Galilei algebra embedded into the dual of the Bargmann algebra. Taking the tensor product of the Poisson bivector associated with the cocycle term yields the $(4,0)$-tensor required to induce the symmetric bracket. Given a generator $\mathcal{F}=H+S$ with $H,S:\mathfrak{b}(3)^*\to\mathbb{R}$, with $H$ some Hamiltonian and $S=\frac{1}{2}(\|\mathbf{p}\|^2+\|\mathbf{g}\times\mathbf{p}\|^2)$ (or any other linear combination of functions of $\|\mathbf{p}\|^2$ and $\|\mathbf{g}\times\mathbf{p}\|^2$), we obtain
\begin{equation}\label{eq:irrevgaleqs}
    \begin{aligned}
    \frac{\boldsymbol{d}}{\boldsymbol{d}\tau}f &= \{H,f\} + ((S,f))\\
    &= \boldsymbol{\zeta} \cdot \left( \frac{\partial H}{\partial \boldsymbol{\zeta}} \times \frac{\partial f}{\partial \boldsymbol{\zeta}} \right) + \mathbf{g} \cdot \left( \frac{\partial H}{\partial \boldsymbol{\zeta}} \times \frac{\partial f}{\partial \mathbf{g}} - \frac{\partial f}{\partial \boldsymbol{\zeta}} \times \frac{\partial H}{\partial \mathbf{g}} \right)  + \mathbf{p} \cdot \left( \frac{\partial H}{\partial \boldsymbol{\zeta}} \times \frac{\partial f}{\partial \mathbf{p}} - \frac{\partial f}{\partial \boldsymbol{\zeta}} \times \frac{\partial H}{\partial \mathbf{p}} \right) \\
&\quad + \mathbf{p} \cdot \left( \frac{\partial H}{\partial \mathbf{g}} \frac{\partial f}{\partial E} - \frac{\partial f}{\partial \mathbf{g}} \frac{\partial H}{\partial E} \right) + M^2\left(\Big(\frac{\partial S}{\partial\mathbf{g}}\cdot\frac{\partial H}{\partial\mathbf{p}}-\frac{\partial H}{\partial\mathbf{g}}\cdot \frac{\partial S}{\partial\mathbf{p}}\Big)\Big(\frac{\partial f}{\partial\mathbf{g}}\cdot\frac{\partial H}{\partial\mathbf{p}}-\frac{\partial H}{\partial\mathbf{g}}\cdot \frac{\partial f}{\partial\mathbf{p}}\Big)\right),
    \end{aligned}
\end{equation}
where $\tau$ parametrises the evolution. This system satisfies the axioms of irreversible mechanics by construction. 

For irreversible dynamics on $\mathfrak{se}(2)^*$, we were able to explicitly construct a generic 2-cocycle by using a vector calculus proxy. For $\mathfrak{sgal}(3)^*$, this is a much more demanding task. A bivector on a 10-dimensional space has 45 independent components. The Schouten-Nijenhuis bracket produces a 3-vector, which has 120 independent components. Therefore, to find a general cocycle explicitly, we would have to write an ansatz for 45 unknown functions and solve a coupled system of 120 linear partial differential equations. While this can likely be solved with computer algebra tools, this is beyond the scope of the present work.

In the next section, we discuss infinite-dimensional irreversible dynamics on the dual of the Lie algebra of vector fields on the circle.

\section{Irreversible dynamics on $\mathfrak{vect}(S^1)^*$}\label{sec:irrevvects1}
In this section, we consider the infinite-dimensional example of irreversible dynamics on $\mathfrak{vect}(S^1)^*$, the dual of the Lie algebra of vector fields over the circle, for which an explicit 2-cocycle is known. This 2-cocycle is the Gel'fand-Fuchs cocycle and is well-known in the context of integrable systems, see for instance \cite{khesin2009geometry} for a detailed discussion of infinite-dimensional groups. We focus on $\mathrm{Diff}^+(S^1)$, the group of orientation-preserving diffeomorphisms over the circle, which is discussed in detail in \cite[Section 2.2]{khesin2009geometry}. The Lie algebra of $\mathrm{Diff}^+(S^1)$ is $\mathfrak{vect}(S^1)$ and the dual of this Lie algebra is the object of interest in this section. 

We fix a coordinate $\theta$ on the circle. Any vector field in $\mathrm{vect}(S^1)$ can be written as $f(\theta)\partial_\theta$ with $f\in C^\infty(S^1)$, where $\partial_\theta$ is shorthand notation for $\frac{\partial}{\partial\theta}$. The Lie bracket on $\mathfrak{vect}(S^1)$ then takes the form
\begin{equation}
[f(\theta)\partial_\theta, g(\theta)\partial_\theta] = \Big(f(\theta)g'(\theta)-g(\theta)f'(\theta)\Big)\partial_\theta.
\end{equation}
Here $f'$ denotes the derivative of $f$ with respect to $\theta$. The smooth dual space $\mathfrak{vect}(S^1)^*$ of the Lie algebra is identified with the space of covector-valued densities $\Omega^1(S^1,T^*S^1)=\{u(\theta)d\theta\otimes d\theta\colon u\in C^\infty(S^1)\}$ with the dual pairing given by
\begin{equation}\label{eq:dualpairingvect}
\langle u(\theta)d\theta\otimes d\theta, f(\theta)\partial_\theta\rangle = \int_{S^1} \Big(\iota_{f(\theta)\partial_\theta} u(\theta)d\theta\Big)d\theta = \int_{S^1}f(\theta)u(\theta)d\theta.
\end{equation}
This allows us to compute the coadjoint action $\mathrm{ad}^*:\mathfrak{vect}(S^1)\times \mathfrak{vect}(S^1)^*\to\mathfrak{vect}(S^1)^*$ and construct the Lie-Poisson bracket. The coadjoint action is given by
\begin{equation}\label{eq:coadjointactionvect}
\mathrm{ad}^*_{f\partial_\theta} u\,d\theta\otimes d\theta = \big((uf)' + uf'\big)d\theta\otimes d\theta.
\end{equation}
To write out the Lie-Poisson bracket in the infinite-dimensional setting, we require the variational (or functional) derivative. While this derivative is well-known, for completeness, we include the definition. For a locally convex topological vector space $X$, consider an open subset $U\subseteq X$ and a functional $F:U\to \mathbb{R}$. The Gateaux differential $\delta F(x;v)$ of $F$ at $x\in U$ in the direction of $v\in X$ is defined as
\begin{equation}
\delta F(x;v)= \lim_{t\to 0} \frac{F(x+tv)-F(x)}{t} = \frac{d}{dt}\bigg|_{t=0}F(x+tv).
\end{equation}
In combination with a weakly non-degenerate dual pairing $\langle\,\cdot\,,\,\cdot\,\rangle:X\times X^*\to\mathbb{R}$, we can obtain the functional derivative $\frac{\delta F}{\delta x}\in X^*$ from this definition. Specifically, taking our space to be $X=\mathfrak{vect}(S^1)^*$ with its predual $X^*=\mathfrak{vect}(S^1)$, if the Gateaux differential of a functional $F: \mathfrak{vect}(S^1)^* \to \mathbb{R}$ is bounded and linear in the variation $v \in \mathfrak{vect}(S^1)^*$, it can be represented by a unique element in the predual via the pairing
\begin{equation}
\delta F(\mu; v) = \left\langle \nu, \frac{\delta F}{\delta \mu} \right\rangle, \quad \forall \nu \in \mathfrak{vect}(S^1)^*.
\end{equation}
The Lie-Poisson bracket on $\mathfrak{vect}(S^1)^*$ for smooth functionals $F,G\in C^\infty(\mathfrak{vect}(S^1)^*)$ is obtained from its general definition as
\begin{equation}\label{eq:liepoissonvect}
\begin{aligned}
\{F,G\}(\mu) &= \left\langle \mu,\left[\frac{\delta F}{\delta \mu}, \frac{\delta G}{\delta \mu}\right]\right\rangle.
\end{aligned}
\end{equation}
Note that each $\mu\in\mathfrak{vect}(S^1)^*$ can be represented by $u(\theta)d\theta\otimes d\theta$ for some choice of $u\in C^\infty(S^1)$. The dual of adjoint action (determined by the Lie bracket) is the coadjoint action \eqref{eq:coadjointactionvect}. This gives us a simple procedure to translate the abstract Lie-Poisson bracket into concrete equations without needing to introduce additional notation to separate formulations using $\mu$ and formulations using the coefficient function $u(\theta)$.

We can now define Hamiltonian dynamics on $\mathfrak{vect}(S^1)^*$ by choosing a Hamiltonian. Common choices are Hamiltonians consisting of purely kinetic energy, typically modelled by metrics, such as $L^2$ and $H^1$, because of their special role in the theory of integrable systems. Consider the $L^2$-metric, which corresponds to the Hamiltonian $H(u(\theta)d\theta\otimes d\theta) = \frac{1}{2}\int_{S^1} u^2 d\theta$. The functional derivative is simply $\frac{\delta H}{\delta u} = u(\theta)\partial_\theta$. The reversible dynamics is obtained from the Lie-Poisson bracket and given in terms of the coadjoint action \eqref{eq:coadjointactionvect} as
\begin{equation}\label{eq:inviscidburgers}
(\partial_t u + 3uu')d\theta\otimes d\theta = 0,
\end{equation}
which one recognises as the inviscid Burgers' equation (or Hopf equation). If one instead chooses the $H^1$-metric $H(u(\theta)d\theta\otimes d\theta) = \frac{1}{2}\int_{S^1} (u^2 + (u')^2) d\theta$, the resulting dynamics yield the Camassa-Holm equation. See \cite[Table 4.1]{khesin2009geometry} for more examples.

The Casimirs of the Lie-Poisson bracket \eqref{eq:liepoissonvect} are much more subtle to characterise than those of finite dimensional Lie-Poisson brackets. If the function $u(\theta)>0$ on the circle, the Lie-Poisson bracket \eqref{eq:liepoissonvect} has a single Casimir, given by
\begin{equation}
S(u(\theta)d\theta\otimes d\theta) = \int_{S^1} \sqrt{u(\theta)}d\theta.
\end{equation}
If $u(\theta)$ changes sign on the circle, then the integrals
\begin{equation}
\int_a^b \sqrt{|u(\theta)|}d\theta,
\end{equation}
evaluated between any two consecutive zeros $a$ and $b$ of the function $u(\theta)$ are invariant. In \cite[Page 72]{khesin2009geometry}, it is shown that for a function $u^\epsilon$ which is everywhere positive for $\epsilon>0$, has a double zero for $\epsilon=0$ and two simple zeros for $\epsilon<0$, the codimension for the coadjoint orbit of $u^\epsilon(\theta)d\theta\otimes d\theta$ changes from 1 for $\epsilon>0$ to 2 for $\epsilon\leq 0$, since the number of Casimirs jumps from 1 to 2. The central extension of $\mathfrak{vect}(S^1)$ remedies this pathological behaviour.

To construct irreversible dynamics on this space, we use the specific 2-cocycle associated with the central extension. The 2-cocycle $\omega:\mathfrak{vect}(S^1)\times\mathfrak{vect}(S^1)\to\mathbb{R}$ given explicitly by
\begin{equation}
\omega\Big(f(\theta)\partial_\theta,g(\theta)\partial_\theta\Big) = \int_{S^1} f'(\theta)g''(\theta)d\theta,
\end{equation}
is the Gel'fand-Fuchs 2-cocycle. The Virasoro algebra $(\mathfrak{vir},[\,\cdot\,,\,\cdot\,]_{\mathfrak{vir}}) = \big(\mathfrak{vect}(S^1)\oplus\mathbb{R},([\,\cdot\,,\,\cdot\,],\omega(\,\cdot\,,\,\cdot\,)\big)$ is the central extension of the algebra of vector fields on the circle, with the Lie bracket explicitly given by
\begin{equation}
[(f\partial_\theta,a),(g\partial_\theta,b)]_\mathfrak{vir} = ([f\partial_\theta,g\partial_\theta],\omega(f\partial_\theta,g\partial_\theta)).
\end{equation}
This cocycle introduces a constant, coordinate-independent bivector term to the Lie-Poisson bracket. For functionals $F$ and $G$ on the dual of the Virasoro algebra, $\mathfrak{vir}^* \cong \mathfrak{vect}(S^1)^* \oplus \mathbb{R}$, the centrally extended Poisson bracket is
\begin{equation}
\{F,G\}_{\mathfrak{vir}}(u(\theta)d\theta\otimes d\theta, c) = \left\langle u(\theta)d\theta\otimes d\theta, \left[ \frac{\delta F}{\delta u}, \frac{\delta G}{\delta u} \right]\right\rangle + c\, \omega\left(\frac{\delta F}{\delta u},\frac{\delta G}{\delta u}\right),
\end{equation}
with the dual pairing given by \eqref{eq:dualpairingvect}. The coadjoint orbits of the centrally extended bracket do not suffer from the codimension parity pathology. In the reversible setting, taking the $L^2$-metric as the Hamiltonian with this extended bracket yields the Korteweg-de Vries equation.

Following the geometric construction as in the finite-dimensional central extensions, we formulate the symmetric bracket as
\begin{equation}
((S, f)) = \omega\left(\frac{\delta S}{\delta u}, \frac{\delta H}{\delta u}\right) \omega\left(\frac{\delta f}{\delta u}, \frac{\delta H}{\delta u}\right),
\end{equation}
It is immediately clear that this construction satisfies the axioms of irreversible mechanics. Because $\omega$ is a $L^2$-skew-symmetric bilinear form, setting $f=H$ makes the second term vanish, guaranteeing energy conservation.

We now choose the $L^2$-metric for the Hamiltonian, $S(u(\theta)d\theta\otimes d\theta) = \int_{S^1}\sqrt{u(\theta)}d\theta$ as the entropy and assume everywhere positive initial data so that this choice of entropy is indeed the Casimir. The variational derivative of the Hamiltonian is then simply $\frac{\delta H}{\delta u}=u\partial_\theta$ and for the entropy we have $\frac{\delta S}{\delta u}=\frac{1}{2\sqrt{u}}\partial_\theta$. The Lie-Poisson bracket \eqref{eq:liepoissonvect} on $\mathfrak{vect}(S^1)^*$ produces the inviscid Burgers' equation \eqref{eq:inviscidburgers}. 

For the irreversible symmetric bracket, the first term evaluates to a global, scalar-valued functional of the state $u$, which we can write in several equivalent ways 
\begin{equation}\label{eq:computation}
\omega\left(\frac{\delta S}{\delta u}, \frac{\delta H}{\delta u}\right) = \int_{S^1} \left(\frac{1}{2\sqrt{u}}\right)' u'' d\theta = -\int_{S^1} \frac{u' u''}{4 u^{3/2}} d\theta = -\int_{S^1} \frac{3(u')^3}{16u^{5/2}}d\theta = -\int_{S^1} \frac{u'''}{2\sqrt{u}}d\theta.
\end{equation}
The third expression in \eqref{eq:computation} is obtained by evaluating the spatial derivative of the variational derivative of the entropy, and the remaining two expressions are obtained by integration by parts. The second term in the symmetric bracket can be integrated by parts to isolate $\frac{\delta f}{\delta u}$. We then obtain
\begin{equation}
\omega\left(\frac{\delta f}{\delta u}, \frac{\delta H}{\delta u}\right) = \int_{S^1} \left(\frac{\delta f}{\delta u}\right)' u'' d\theta = - \int_{S^1} \frac{\delta f}{\delta u} u''' d\theta.
\end{equation}
We can now assemble the total evolution of an observable $f$. The irreversible equation of motion on $\mathfrak{vect}(S^1)^*$ is given by
\begin{equation}
\frac{\boldsymbol{d}}{\boldsymbol{d}t} f = \int_{S^1} \frac{\delta f}{\delta u} \Bigg( -3uu' - \Big(\int_{S^1}\frac{u'''}{2\sqrt{u}}d\theta\Big) u''' \Bigg) d\theta.
\end{equation}
Because this relation must hold for any arbitrary observable functional $f$, we can extract the strong form of the partial differential equation governing the underlying state $u(\theta, t)$. The irreversible dynamics of the state is governed by
\begin{equation}
\left(\frac{\partial u}{\partial t} + 3uu' + \Big(\int_{S^1}\frac{u'''}{2\sqrt{u}} d\theta\Big) u'''\right)d\theta\otimes d\theta = 0.
\end{equation}
The geometric insertion of entropy production via the Gel'fand-Fuchs 2-cocycle manifests as a dynamically modulated dispersion coefficient in the Korteweg-de Vries equation. The dispersion coefficient itself is interesting, as it is the ratio of the dispersion term and the variational derivative of the entropy function. Furthermore, this coefficient is nonlocal and depends on the solution in a highly nonlinear fashion. By construction, the entropy is increasing, but at the same time, the $L^2$-norm of the solution is preserved. 

\section{Discussion and conclusion}\label{sec:conclusion}
In this work, we introduced an explicit method for constructing irreversible dynamics on Poisson manifolds. This method relies solely on the intrinsic Poisson structure, its kernel, and the existence of nontrivial Lichnerowicz-Poisson 2-cocycles. Using these geometric ingredients, we built a symmetric bracket that satisfies the axioms of metriplectic mechanics and the GENERIC framework that does not depend on the choice of a Riemannian metric.

We demonstrated the method on three Lie algebra duals: $\mathfrak{se}(2)^*$ (the special Euclidean group), $\mathfrak{sgal}(3)^*$ (the special Galilei group), and $\mathfrak{vect}(S^1)^*$ (the infinite-dimensional group of orientation-preserving diffeomorphisms of the circle). While the theory is not restricted to Lie algebra duals and one may consider more general Poisson manifolds, the linear setting is especially convenient for explicit computations.

For the dual of the special Euclidean algebra, we worked with two distinct Lichnerowicz-Poisson 2-cocycles. The first arose from a central extension, yielding a constant Poisson bivector. The second 2-cocycle was constructed directly using 3D vector proxies. We showed that this generic 2-cocycle violates the Jacobi identity and therefore does not yield a valid Poisson bracket on its own. Nevertheless, both options successfully generate irreversible dynamics that satisfy the axioms of metriplectic mechanics and GENERIC. This demonstrates that the 2-cocycle driving the dissipation does not need to be a Poisson bracket itself.

We then considered the dual of the Galilei Lie algebra. The Lie-Poisson bracket in this case possesses a richer kernel than that of the special Euclidean algebra, which we explicitly characterised via the method of characteristics, allowing for a broader class of admissible entropy functionals.

Finally, we provided an infinite-dimensional example by studying the construction on the dual of the Lie algebra of vector fields over the circle. When taking the $L^2$-metric as the Hamiltonian, the resulting irreversible dynamics manifest as a Korteweg-de Vries equation with a dynamically modulated dispersion coefficient that is non-local and depends on the solution in a highly nonlinear fashion. While one can employ standard ODE techniques and Lyapunov-based arguments to address local and global well-posedness for the finite-dimensional examples, establishing well-posedness for this novel infinite-dimensional integro-differential equation will require sophisticated PDE techniques, opening an interesting avenue for future analytical work.

\section*{Acknowledgements}
EL expresses gratitude towards Sonja Cox, Darryl Holm, Ruiao Hu, Bas Janssens, Eric Opdam and Oliver Street for many insightful discussions during the preparation of this work. Several of the anonymous referees made valuable comments to improve this work. EL was supported by NWO grant VI.Vidi.213.070.

\section*{Conflict of Interest Statement}
The author has no competing interests to declare that are relevant to the content of this work. The author certifies that
they have no affiliations with or involvement in any organization or entity with any financial interest or non-financial
interest in the subject matter or materials discussed in this manuscript.

\section*{Data Availability Statement}
This article does not contain any data.

\bibliographystyle{plainnat}
\bibliography{biblio}

\end{document}